\documentclass[runningheads]{llncs}

\usepackage{graphicx}
\usepackage{cite}
\usepackage{amsmath}
\usepackage{amsfonts}
\usepackage{amssymb}
\usepackage{multirow}
\usepackage[hidelinks]{hyperref}
\usepackage{algorithm}
\usepackage{algpseudocode}
\usepackage{url}
\usepackage{subcaption}
\usepackage{mathtools}
\usepackage[title]{appendix}

\usepackage{xcolor}

\usepackage{environ}
\usepackage{setspace} 

\newenvironment{alignSmall}{\nobreak\small\noindent\align}{\endalign}
\newenvironment{alignFootnotesize}{\nobreak\footnotesize\noindent\align}{\endalign}
\newenvironment{alignScriptsize}{\nobreak\scriptsize\noindent\align}{\endalign}

\NewEnviron{myAlignS}{%
	\begin{singlespace}
		\vspace{-3pt}
		\begin{alignSmall}
			\BODY
		\end{alignSmall}
		\vspace{-5pt}
	\end{singlespace}
}

\NewEnviron{myAlignSS}{%
	\begin{singlespace} 
		\vspace{-3pt}
		\begin{alignFootnotesize}
			\BODY
		\end{alignFootnotesize}
		\vspace{-5pt}
	\end{singlespace}
}

\NewEnviron{myAlignSSS}{%
	\begin{singlespace}
		\vspace{-5pt}
		\begin{alignScriptsize}
			\BODY
		\end{alignScriptsize}
		\vspace{-5pt}
	\end{singlespace}
}

\begin{document}
	
\graphicspath{{figures/}}

%
\title{PGLP: Customizable and Rigorous Location Privacy through Policy Graph \thanks{This work is partially supported by JSPS KAKENHI Grant No. 17H06099, 18H04093, 19K20269, U.S. National Science Foundation (NSF) under CNS-2027783 and CNS-1618932, and Microsoft Research Asia (CORE16).}}
%
%
\author{Yang Cao \inst{1} \and
Yonghui Xiao \inst{2} \thanks{Yang and Yonghui contributed equally to this work. } \and
Shun Takagi \inst{1} \and
Li Xiong \inst{2}  \and
Masatoshi Yoshikawa \inst{1}  \and
Yilin Shen \inst{3} \and
Jinfei Liu  \inst{2}  \and
Hongxia Jin  \inst{3} \and
Xiaofeng Xu \inst{2}
}
\authorrunning{Y. Cao and Y. Xiao et al.}
%
\institute{Kyoto University  \\  \email{yang@i.kyoto-u.ac.jp,  s.takagi@db.soc.i.kyoto-u.ac.jp, yoshikawa@i.kyoto-u.ac.jp} \and
Emory University \\ \email{\{lxiong, jliu253\}@emory.edu \{yohuxiao, xuxiaofeng1989\}@gmail.com} \and
Samsung Research America \\ \email{\{yilin.shen, hongxia.jin\}@samsung.com} 
}

\maketitle              


\begin{abstract}
Location privacy has been extensively studied in the literature.
However, existing location privacy models are either not rigorous or not customizable, which limits the trade-off between privacy and utility in many real-world applications.
To address this issue, we propose a new location privacy notion called PGLP, i.e., \textit{Policy Graph based Location Privacy},  providing a rich interface to release private locations with customizable and rigorous privacy guarantee.
First, we design a rigorous privacy  for PGLP by extending differential privacy.
Specifically, we formalize location privacy requirements using a \textit{location policy graph}, which is expressive and customizable.
Second, we investigate how to satisfy an arbitrarily given location policy graph under realistic adversarial knowledge, which can be seen as constraints or public knowledge about user's mobility pattern.
We find that a  policy graph may not always be viable and may suffer \textit{location exposure} when the attacker knows the user's mobility pattern.
We propose efficient methods to detect location exposure and repair the policy graph with optimal utility.
Third, we design an end-to-end  location trace release framework  that pipelines the detection of location exposure,  policy graph repair, and private location release at each timestamp with customizable and rigorous location privacy.
Finally,  we conduct experiments on real-world datasets to verify the effectiveness  and the efficiency of the proposed algorithms.
\end{abstract}

\vspace{-25pt}
\section{Introduction}
\label{sec-intro}
\vspace{-5pt}
As GPS-enabled devices such as smartphones or wearable gadgets are pervasively used and rapidly developed, location data have been continuously generated, collected, and analyzed.
These personal location data connecting the online and offline worlds are precious, 
because they could be of great value for the society to enable  ride sharing, traffic management, emergency planning, and disease outbreak control as in the current covid-19 pandemic via contact tracing, disease spread modeling, traffic and social distancing monitoring \cite{luo_deepeye:_2020, furuhata_ridesharing:_2013, cao_panda:_2020, ingle_slowing}.   


On the other hand, privacy concerns hinder the extensive use of big location data generated by users in the real world.
Studies have shown that location data could reveal sensitive personal information such as home and workplace, religious and sexual inclinations  \cite{recabarren_what_2017}.
According to a survey \cite{Fawaz:2014:CCS}, $78\%$ smartphone users among $180$ participants believe that Apps accessing their location pose privacy threats.
As a result, the study of \textit{private location release} has drawn increasing research interest and
many location privacy models have been proposed in the last decades (see survey \cite{primault_long_2018}).

However, existing location privacy models for private location releases are either not rigorous or not customizable.
Following the seminal paper \cite{gruteser_anonymous_2003}, the early location privacy models  were designed based on $ k $-anonymity\cite{sweeney_k-anonymity:_2002} and adapted to different scenarios such as  mobile P2P environments \cite{chow_spatial_2011}, trajectory release \cite{bettini_protecting_2005} and personalized $k$-anonymity for location privacy \cite{gedik_protecting_2008}.
The follow-up studies revealed that $ k $-anonymity might not be rigorous because it syntactically defines privacy as a property of the final ``anonymized'' dataset \cite{li_differential_2016} and thus suffers many realistic attacks when the adversary has background knowledge about the dataset \cite{li_t-closeness:_2007, machanavajjhala_l-diversity:_2006}.
To this end, the state-of-the-art location privacy models\cite{geoi_ccs13, xiao_ccs15, chatzikokolakis_constructing_2015, takagi_GGI_2019} were extended from differential privacy (DP) \cite{Dwork06differentialprivacy}  to private location release since DP is considered a rigorous privacy notion which defines privacy as a property of the algorithm.
Although these DP-based location privacy models are rigorously defined,  they are not customizable for different scenarios with various requirements on {privacy-utility trade-off}. 
Taking an example of Geo-Indistinguishability\cite{geoi_ccs13}, which is the first DP-based location privacy, the protection level is  solely controlled by a parameter $ \epsilon $ to achieve indistinguishability between any two possible locations (the indistinguishability is scaled to the Euclidean distance between any two possible locations).

This one-size-fits-all approach may not fit every application's requirement on utility-privacy trade-off.
Different location-based services (LBS) may have different usage of the data and thus need different \textit{location privacy policies} to strike the right balance between privacy and utility.
For instance, a proper location privacy policy for weather apps could be ``\textit{allowing the app to access a user's city-level location but ensuring indistinguishability among locations in each city}'', which guarantees both reasonable privacy and high usability for a city-level weather forecast.
Similarly, for POI recommendation \cite{bao_recommendations_2015}, trajectory mining \cite{parent_semantic_2013} or crowd monitoring during the pandemic \cite{ingle_slowing}, a suitable location privacy policy could be ``\textit{allowing the app to access the semantic category (e.g., a restaurant or a shop) of a user's location but ensuring indistinguishability among locations with the same category}'', so that the LBS provider may know the user is at a restaurant or a shop, but not sure which restaurant or which shop.

In this work, we study \textit{how to release private  location with customizable and rigorous privacy}.
There are three significant challenges to achieve this goal.
First, there is a lack of a rigorous and customizable location privacy metric and mechanisms.
The closest work regarding customizable privacy is Blowfish privacy \cite{Blowfish-SIGMOD14} for statistical data release, which uses a graph to represent a customizable privacy requirement, in which a node indicates a possible database instance to be protected, and an edge represents indistinguishability between the two possible databases.
Blowfish privacy and its mechanisms are not applicable in our setting of private location release.
It is because Blowfish privacy is defined on a statistical query over database with multiple users' data; whereas the input in the scenario of private location release is a single user's location.

The second challenge is how to satisfy an arbitrarily given location privacy policy under realistic adversarial knowledge, which is public knowledge about users' mobility pattern.
In practice, as shown in \cite{xiao_ccs15, xiao_loclok:_2017}, an adversary could take advantage of side information to rule out inadmissible locations\footnote{For example, it is impossible to move from Kyoto to London in a short time.} and reduce a user's possible locations into a small set, which we call \textit{constrained domain}.
 We find that the location privacy policy may not be viable under a constrained domain and the user may suffer location exposure (we will elaborate how this could happen in Sec. \ref{sec-node-exposure}).

\begin{figure}[t]
	\centering
	\begin{subfigure}{0.55\textwidth}
		\includegraphics[width=6cm]{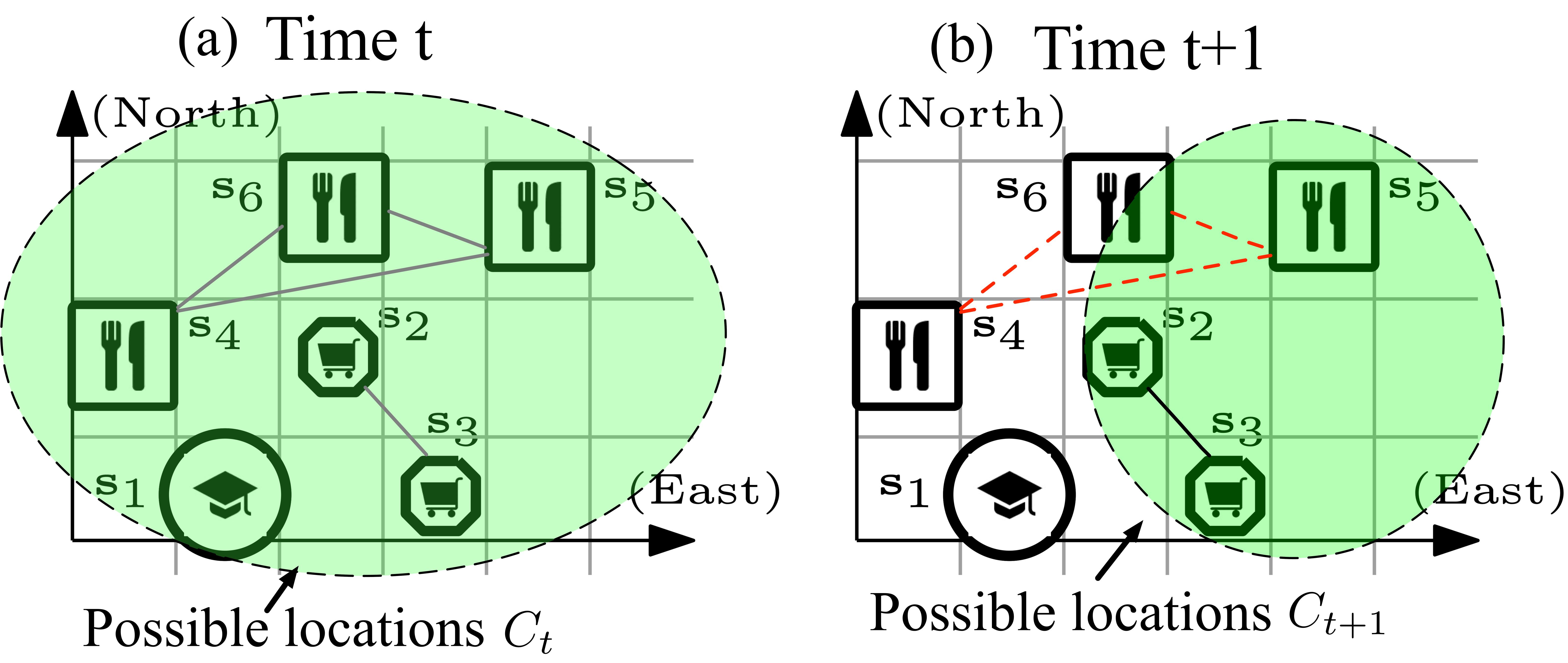}
	\end{subfigure}
\vspace{-5pt}
	\caption{An example of location policy graph and the constrained domains (i.e., possible locations) $ C_t $ and $ C_{t+1} $ at time $ t $ and $ t+1 $, respectively.}
	\label{fig:ex1}
	\vspace{-18pt}
\end{figure}

The third challenge is how to release private locations continuously with high utility.
We attempt to provide an end-to-end solution that takes the user's true location and a predefined location privacy policy as inputs, and outputs private location trace on the fly. 
 We summarize the research questions  below.
\vspace{-5pt}
\begin{itemize}
	\item How to design a rigorous location privacy metric with customizable location privacy policy? (Section \ref{sec-pglp})
	\item How to detect the problematic location privacy policy and repair it with high utility? (Sections \ref{sec-exposure}) 
	\item How to design an end-to-end framework to release private location continuously? (Section \ref{sec-release})
\end{itemize}

\vspace{-10pt}
\subsection{Contributions}

In this work, we propose a new location privacy metric and mechanisms for releasing private location trace with flexible and rigorous privacy.
To the best of our knowledge, this is the first  DP-based location privacy notion with customizable privacy.
Our contributions are summarized below.

First, we formalize {P}olicy {G}raph based {L}ocation {P}rivacy (PGLP),  which is a rigorous privacy metric extending differential privacy with a customizable \textit{location policy graph}.
Inspired by the statistical privacy notion of Blowfish privacy \cite{Blowfish-SIGMOD14}, we design 
location policy graph to represent which information needs to be protected and which does not.
In a location policy graph (such as the one shown in Fig.\ref{fig:ex1}), the nodes are the user's possible locations, and the edges indicate the privacy requirements regarding the connected locations: an attacker should not be able to significantly distinguish which location is more probable to be the user's true location by observing the released location.
PGLP is a general location privacy model compared with the prior art of DP-based location privacy notions, such as Geo-Indistinguishability \cite{geoi_ccs13} and Location Set Privacy \cite{xiao_ccs15}.
We prove that they are two instances of PGLP  under the specific configurations of the location policy graph.
We also design mechanisms for PGLP by adapting the Laplace mechanism and Planar Isotropic Mechanism (PIM) (i.e., the optimal mechanism for Location Set Privacy  \cite{xiao_ccs15}) w.r.t. a given location policy graph.

Second, we design algorithms that examine the feasibility of a given location policy graph under adversarial knowledge about the user's mobility pattern modeled by Markov Chain. 
We find that the policy graph may not always be viable.
Specially, as shown in Fig.\ref{fig:ex1}, some nodes (locations) in a  policy graph may be  \textit{excluded} (e.g., $ s_4 $ and $s_6 $ in Fig.\ref{fig:ex1} (b)) or \textit{disconnected} (e.g., $ s_5 $ in Fig.\ref{fig:ex1} (b)) due to the limited set of the possible locations.
Protecting the excluded nodes is a lost cause, but it is necessary to protect the disconnected nodes since it may lead to location exposure when the user is at such a location. 
Surprisingly, we find that a disconnected node may \textit{not always}  result in the location exposure, which also depends on the protection strength of the mechanism.
Intuitively, this happens when a mechanism ``overprotects'' a location policy graph by implicitly guaranteeing indistinguishability that is not enforced by the policy.
We design an algorithm to detect the disconnected nodes that suffer location exposure , which are named \textit{isolated node}.
We also design a \textit{graph repair} algorithm to ensure no isolated node in a location policy graph by adding an optimal edge between the isolated node and another node with high utility.

Third, we propose an end-to-end private location trace release framework with PGLP that takes inputs of the user's true location at each time $ t $ and outputs private location continuously satisfying a pre-defined location policy graph. 
The framework pipelines the calculation of constrained domains, isolated node detection, policy graph repair, and private location release mechanism. 
We also reason about the overall privacy guarantee in multiple releases.

Finally, we implement and evaluate the proposed algorithms on real-world datasets, showing that privacy and utility can be better tuned with customizable location policy graphs.

\vspace{-10pt}

\section{Preliminaries}

\vspace{-5pt}
\subsection{Location Data Model}
\label{sec-loc-model}
Similar to \cite{xiao_ccs15,xiao_loclok:_2017}, we employ two coordinate systems to represent locations for applicability for different application scenarios.
A location can be represented by an index of \textit{grid coordinates} or by a two-dimension vector of \textit{longitude-latitude coordinate} to indicate any location on the map.
Specifically, we partition the map into a grid such that each grid cell corresponds to an area (or a point of interest); any location represented by a longitude-latitude coordinate will also have a grid number or index on the grid coordinate. 
We denote the location domain as
$\mathcal{S}=\{\textbf{s}_1,\textbf{s}_2,\cdots, \textbf{s}_N\}$ where each $\textbf{s}_i$ corresponds to a grid cell on the map, $1\leq i \leq N$. 
We use $\textbf{s}_t^*$  and $\textbf{z}_t$ to denote the user's true location and perturbed location at time $ t $. 
We also use $ t $ in $\textbf{s}^*$  and $\textbf{z}$  to refer the locations at a single time when it is clear from the context.

\noindent{\bf Location Query}
For the ease of reasoning about privacy and utility, we use a location query $f: \mathcal{S}\rightarrow \mathbb{R}^2$ to represent the mapping from locations to the longitude and latitude of the center of the corresponding grid cell. 

\vspace{-10pt}
\subsection{Problem Statement}
\label{sec-prob-state}

\vspace{-5pt}

Given a moving user on a map $\mathcal{S}$ in a time period $\{1,2,\cdots,T\}$, our goal is to release the perturbed locations of the user to untrusted third parties at each timestamp under a pre-defined location privacy policy.
We define $ \epsilon $-Indistinguishability as a building block for reasoning about our privacy goal.

\begin{definition}[$\epsilon$-Indistinguishability]
	\label{def-eps-ind}
	Two locations $\textbf{s}_{i}$ and $\textbf{s}_{j}$ are $\epsilon$-indistin-guishable under a randomized mechanism $\mathcal{A}$ iff for any output $\textbf{z} \subseteq Range(\mathcal{A})$, we have $
	\frac{\Pr(\mathcal{A}(\textbf{s}_i)=\textbf{z} )}{\Pr(\mathcal{A}(\textbf{s}_j)=\textbf{z} )}\leq e^{\epsilon}
	$, where $ \epsilon \geq 0 $.
\end{definition}

As we exemplified in the introduction, different LBS applications may have different metrics of utility.
We aim at providing better utility-privacy trade-off by customizable $ \epsilon $-Indistinguishability between locations.


\vspace{-10pt}

\subsubsection{Adversarial Model}
\vspace{-5pt}
We assume that the attackers know the user's mobility pattern modeled by Markov chain, which is widely used for modeling user mobility profiles \cite{gambs2012next, cao_quantifying_2019}.
We use matrix $\textbf{M}\in [0,1]^{N\times N}$ to denote the transition probabilities of Markov chain with $m_{ij}$ being the probability of moving from location $\textbf{s}_i$ to location $\textbf{s}_j$. 
Another adversarial knowledge  is the initial probability distribution of the user's location at $t=1$.
To generalize the notation, we denote probability distribution of the user's location at $t$ by a vector $\textbf{p}_t\in [0,1]^{1\times N}$, and denote the $i$th element in $\textbf{p}_t$ by $ \textbf{p}_t[i]=\Pr(\textbf{s}^*_t=\textbf{s}_i) $,
where $ \textbf{s}^*_t $ is the user's true location at $ t $ and $\textbf{s}_i\in\mathcal{S}$.
Given the above knowledge, the attackers could infer the user's possible locations at time $ t $, which is probably smaller than the location domain $\mathcal{S}$, and we call it a \textit{constrained domain}.

\begin{definition} [Constrained domain]
\label{def-constrained-domain}
We denote {\small $ \mathcal{C}_t = \{\textbf{s}_i |  \Pr(\textbf{s}^*_t=\textbf{s}_i) >0, \textbf{s}_i\in \mathcal{S} \}$} as constrained domain, which indicates a set of possible locations at $ t $.
\end{definition}

We note that the constrained domain can be explained as the requirement of LBS applications. 
For example, an App  could only be  used within a certain area, such as a university free shuttle tracker.

\vspace{-10pt}
\section{Policy Graph based Location Privacy}
\label{sec-pglp}
\vspace{-8pt}

In this section, we first formalize the privacy requirement using \textit{location policy graph} in Sec. \ref{subsec-lpg}.
We then design the privacy metric of PGLP in Sec. \ref{subsec-metric}.
Finally, we propose two mechanisms for PLGP  in Sec. \ref{subsec-mech}.

\vspace{-3pt}

\subsection{Location Policy Graph}
\label{subsec-lpg}

Inspired by Blowfish privacy\cite{Blowfish-SIGMOD14}, we use an undirected graph to define what should be protected, i.e., privacy policies.
The nodes are secrets, and the edges are the required indistinguishability, which indicates an attacker should not be able to distinguish the input secrets by observing the perturbed output.  
In our setting, we treat possible locations as nodes and the indistinguishability between the locations as edges.

\begin{definition}[Location Policy Graph]
	A location policy graph is an undirected graph {\small $\mathcal{G}=(\mathcal{S},\mathcal{E})$} where
	$\mathcal{S}$ denotes all the locations (nodes) and $\mathcal{E}$ represents indistinguishability (edges) between these locations.
\end{definition}

\begin{definition}[Distance in Policy Graph]
We define the distance between two nodes $\textbf{s}_i$ and $\textbf{s}_j$ in a policy graph as the length of the shortest path (i.e., hops) between them, denoted by $d_\mathcal{G}(\textbf{s}_i, \textbf{s}_j)$. 
If    $\textbf{s}_i$ and $\textbf{s}_j$  are disconnected, {\small $d_\mathcal{G}(\textbf{s}_i, \textbf{s}_j)  = \infty$}.
\end{definition}

In DP, the two possible database instances with or without a user's data are called  \textit{neighboring databases}.
In our location privacy setting, we define neighbors as two nodes with an edge in a policy graph.

\begin{definition}[Neighbors]
\label{def-neighbors}
	The neighbors of location $\textbf{s}$, denoted by $\mathcal{N}(\textbf{s})$, is the set of nodes having an edge with $\textbf{s}$, i.e.,	$\mathcal{N}(\textbf{s})= \{\textbf{s}' | d_\mathcal{G}(\textbf{s},\textbf{s}') =  1, \textbf{s}'\in \mathcal{S}\}$.
\end{definition}

We denote the nodes having a path with  $\textbf{s}$ by $\mathcal{N}^P (\textbf{s})$, i.e., the nodes in the same connected component with $ \textbf{s} $.
In our framework, we assume the  policy graph is given and public.
In practice, the location privacy policy can be defined application-wise  and identical for all users using the same application.

\vspace{-10pt}
\subsection{Definition of PGLP}
\label{subsec-metric}

We now formalize Policy Graph based Location Privacy (PGLP), which guarantees $\epsilon$-indistinguishability in Definition \ref{def-eps-ind} for every pair of neighbors (i.e., for each edge) in a given location policy graph.

\vspace{-5pt}
\begin{definition}[$\{\epsilon,\mathcal{G}\}$-Location Privacy]
	\label{def-PGLP}
	A randomized algorithm $\mathcal{A}$  satisfies $\{\epsilon,\mathcal{G}\}$-location privacy iff for all $  \textbf{z} \subseteq Range(\mathcal{A})$ and for all pairs of neighbors $   \textbf{s} $ and $  \textbf{s}'$ in $\mathcal{G}$, we have   {\small $
	\frac{\Pr(\mathcal{A}(\textbf{s})=\textbf{z} )}{\Pr(\mathcal{A}(\textbf{s}')=\textbf{z} )}\leq e^{\epsilon}
$}.
	\vspace{-5pt}
\end{definition}

In PGLP,  privacy is rigorously guaranteed through ensuring indistinguishability between any two neighboring locations specified by a customizable location policy graph.
The above definition implies the indistinguishability between two nodes that have a path in the policy graph.

\begin{lemma}
\label{lm-graph-dist}
An algorithm $\mathcal{A}$ satisfies $\{\epsilon,\mathcal{G}\}$-location privacy, iff any two  nodes $\textbf{s}_i, \textbf{s}_j \in \mathcal{G} $ are $ \epsilon \cdot d_\mathcal{G}(\textbf{s}_i, \textbf{s}_j)$-indistinguishable.
\end{lemma}

\vspace{-5pt}
Lemma \ref{lm-graph-dist} indicates that,  if there is a path between two nodes  $\textbf{s}_i, \textbf{s}_j$ in the policy graph, the corresponding indistinguishability is required at a certain degree; if two nodes are disconnected, the indistinguishability  is not required  (i.e., can be $ \infty $) by the policy.
As an extreme case, if a node is disconnected with any other nodes, it is allowed  to be released  without any perturbation.

\vspace{-15pt}

\subsubsection{Comparison with Other Location Privacy Models.}

We analyze the relation between PGLP and two well-known DP-based location privacy models, i.e., {Geo-Indistinguishability} (Geo-Ind) \cite{geoi_ccs13} and $\delta$-Location Set Privacy \cite{xiao_ccs15}. 
We show that PGLP can represent them under proper configurations of policy graphs.

{\textit{Geo-Ind} \cite{geoi_ccs13}} guarantees a level of indistinguishability between two locations $\textbf{s}_i$ and $\textbf{s}_j$ that is scaled with their Euclidean distance, i.e.,  $\epsilon \cdot d_E(\textbf{s}_i, \textbf{s}_j) $-indistinguish-ability, where $d_E(\cdot, \cdot )$ denotes  Euclidean distance. 
Note that the unit length used in Geo-Ind scales the level of indistinguishability. 
We assume that, for any neighbors  $ \textbf{s}$ and $ \textbf{s}'$, the unit length  used in Geo-Ind makes $ d_E(\textbf{s}, \textbf{s}') \geq 1 $.

\begin{figure}[t]
	\centering
	\includegraphics[width=6cm]{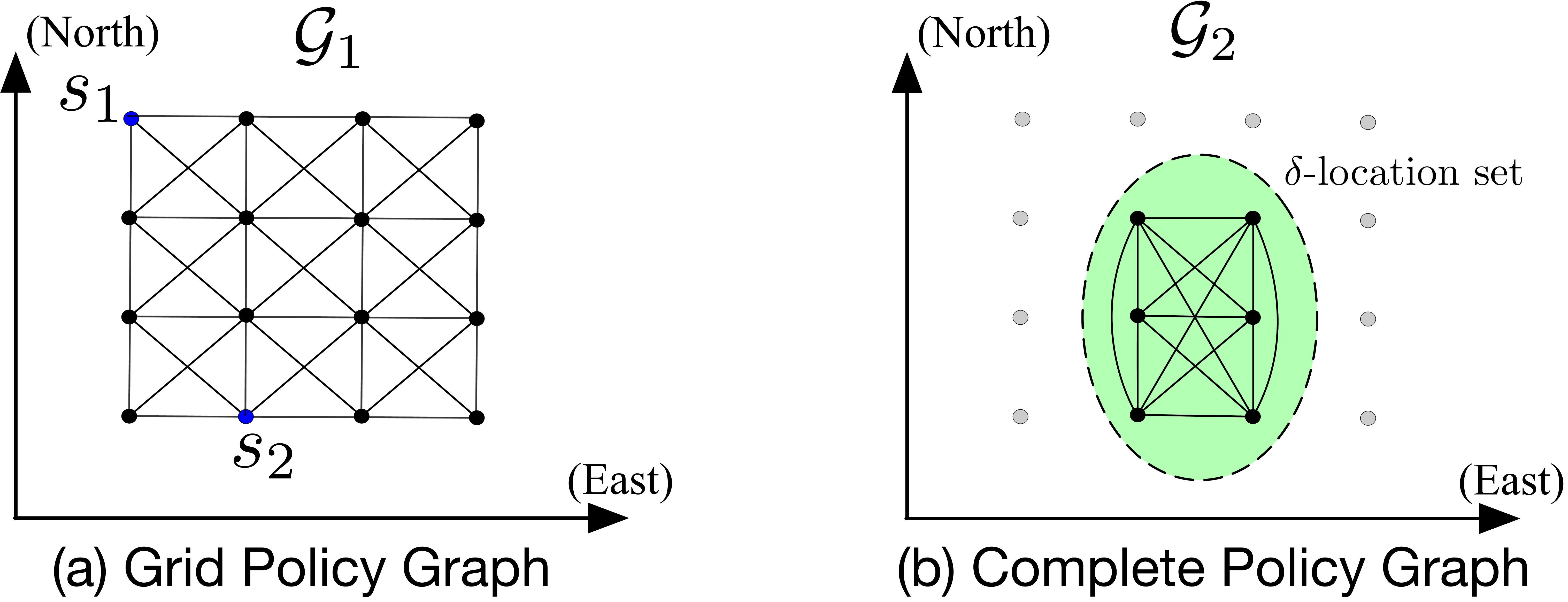}
	\vspace{-5pt}
	\caption{Two examples of location policy graphs.}
	\label{figure-two-graphs}
	\vspace{-20pt}
\end{figure}

Let $\mathcal{G}_1$ be a location policy graph that every location has edges with its closest eight locations on the map, as shown in Fig.\ref{figure-two-graphs} (a). 
We can derive Theorem  \ref{thm:geo-ind} by Lemma \ref{lm-graph-dist} with the fact of {\small $d_\mathcal{G}(\textbf{s}_i, \textbf{s}_j) \leq d_E(\textbf{s}_i, \textbf{s}_j)$} for any {\small $\textbf{s}_i, \textbf{s}_j \in \mathcal{G}_1$} (e.g., in Fig.\ref{figure-two-graphs}(a), {\small $d_\mathcal{G}(\textbf{s}_1, \textbf{s}_2) =3$} and {\small $d_E(\textbf{s}_1, \textbf{s}_2) = \sqrt{10}$}).

\vspace{-5pt}
\begin{theorem}
\label{thm:geo-ind}
An algorithm satisfying $\{\epsilon, \mathcal{G}_1\}$-location privacy also achieves $\epsilon$-Geo-Indistinguishability.
\end{theorem}

\textit{$\delta$-Location Set Privacy} \cite{xiao_ccs15} extends differential privacy on a subset of possible locations, which is assumed as adversarial knowledge.
We note that the constrained domain in Definition \ref{def-constrained-domain} can be considered a generalization of $ \delta $-location set, whereas we do not specify the calculation of this set in PGLP.
$\delta$-Location Set Privacy ensures indistinguishability among any two locations in the $\delta$-location set.
Let $\mathcal{G}_2$ be a location policy graph that is complete, i.e., fully connected among all locations in the $\delta$-location set as shown in Fig.\ref{figure-two-graphs}(b). 

\vspace{-5pt}
\begin{theorem}
	An algorithm satisfying $\{\epsilon, \mathcal{G}_2\}$-location privacy also achieves $\delta$-Location Set privacy. 
\end{theorem}
\vspace{-5pt}

\noindent We defer the proofs of the theorems to a full version because of space limitation.

\vspace{-10pt}
\subsection{Mechanisms for PGLP}
\label{subsec-mech}
\vspace{-5pt}
In the following, we show how to transform existing DP mechanisms into one satisfying PGLP using \textit{graph-calibrated sensitivity}.
We temporarily assume the constrained domain $ \mathcal{C} = \mathcal{S} $ and  study the effect of  $\mathcal{C}$  on policy  $\mathcal{G}$ in Section \ref{sec-exposure}.

As shown in Section \ref{sec-loc-model},  the problem of private location release can be seen as answering a location query {\small {\small $f:\mathcal{S}\rightarrow\mathbb{R}^2$}} privately.
Then we can adapt the existing DP mechanism for releasing private locations by adding random noises to longitude and latitude independently.
We use this approach below to adapt the Laplace mechanism and Planar Isotropic Mechanism (PIM) (i.e., an optimal mechanism for Location Set Privacy \cite{xiao_ccs15}) to achieve PGLP.

\vspace{-8pt}
\subsubsection{\textbf{Policy-based  Laplace Mechanism (P-LM).}}
Laplace mechanism is built on the $\ell_1$-norm sensitivity \cite{Dwork-calibrating}, defined as the maximum change of the query results due to the difference of neighboring databases.
In our setting, we calibrate this sensitivity w.r.t. the neighbors specified in a location policy graph.

\vspace{-5pt}
\begin{definition}[Graph-calibrated $\ell_1$-norm Sensitivity]
	\label{def-l1}
	For a location $\textbf{s}$ and a query {\small $f(\textbf{s})$: $\textbf{s}\rightarrow \mathbb{R}^2$}, its $\ell_1$-norm sensitivity {\small $S^\mathcal{G}_f$} is the maximum $\ell_1$ norm of {\small $\Delta f ^ \mathcal{G}$}
	where {\small $\Delta f ^ \mathcal{G}$} is a set of points (i.e., two-dimension vectors) of
	{\small  $\big(f(\textbf{s}_i)-f(\textbf{s}_j)\big)$}  for  {\small  $\textbf{s}_i, \textbf{s}_j\in \mathcal{N}^P(\textbf{s})$} (i.e., the nodes with the same connected component of $ \textbf{s} $).
\end{definition}

\vspace{-5pt}
			
We note that, for a true location $ \textbf{s} $, releasing $\mathcal{N}^P (\textbf{s})$ does not violate the privacy  defined by the policy graph. 
It is because, for any connected $ \textbf{s} $ and $ \textbf{s}' $, $\mathcal{N}^P (\textbf{s})$ and  $\mathcal{N}^P (\textbf{s}')$ are the same; 
while, for any disconnected $ \textbf{s} $ and $ \textbf{s}' $, the indistinguishability between $\mathcal{N}^P (\textbf{s})$ and  $\mathcal{N}^P (\textbf{s}')$ is not required by Definition \ref{def-PGLP}.

\vspace{-15pt}
\begin{algorithm}[H]
	\scriptsize
	\caption{\small Policy-based  Laplace Mechanism (P-LM)}
	\begin{algorithmic}[1]
		\Require{$ \epsilon $,  ${\mathcal{G}}$,  the user's true location $\textbf{s}$.}
		\State{Calculate  $S_f^\mathcal{G}  = sup|| \big(f(\textbf{s}_i)-f(\textbf{s}_j)\big)||_1  $ for all  neighbors $ \textbf{s}_i, \textbf{s}_j \in\mathcal{N}^P(\textbf{s}) $;}
		\State{Perturb location $\textbf{z}' = f(\textbf{s})+ [Lap(S_f^\mathcal{G} /\epsilon), Lap(S_f^\mathcal{G} /\epsilon)]^T$;}
		\State{\textbf{return} a location $\textbf{z} \in  \mathcal{S}$ that is closest to $\textbf{z}'$  on the map.}
	\end{algorithmic}
	\label{algo-p-lm}
\end{algorithm}
	\vspace{-20pt}


\vspace{-5pt}
\begin{theorem}
\label{thm1}
P-LM satisfies $\{\epsilon, \mathcal{G}\}$-location privacy.
\end{theorem}


\vspace{-15pt}
	
\subsubsection{\textbf{Policy-based Planar Isotropic Mechanism (P-PIM).}}
PIM \cite{xiao_ccs15}  achieves the low bound of differential privacy on two-dimension space for Location Set Privacy.
It adds noises to longitude and latitude using $K$-norm mechanism \cite{Geometry-Hardt-STOC10} with  \textit{sensitivity hull} \cite{xiao_ccs15}, which extends the convex hull of the sensitivity space in $K$-norm mechanism.
We propose a \textit{graph-calibrated sensitivity hull}  for PGLP.

\vspace{-5pt}
\begin{definition}[Graph-calibrated Sensitivity Hull]
	\label{def-hull}
	For  a location $\textbf{s}$ and a query $f(\textbf{s})$: $\textbf{s}\rightarrow \mathbb{R}^2$, the graph-calibrated sensitivity hull $K(\mathcal{G})$ is the convex hull of $\Delta f ^ \mathcal{G}$
	where $\Delta f ^ \mathcal{G}$ is a set of points (i.e., two-dimension vectors) of
	$\big(f(\textbf{s}_i)-f(\textbf{s}_j)\big)$ for any {\small  $\textbf{s}_i, \textbf{s}_j\in \mathcal{N}^P(\textbf{s})$} and $\textbf{s}_i, \textbf{s}_j$ are neighbors, i.e., {\small $K(\mathcal{G}) = Conv(\Delta f ^ \mathcal{G}) $}. 
\end{definition}

We note that, in Definitions \ref{def-l1} and \ref{def-hull}, the sensitivities are independent of the true location $ \textbf{s} $ and  all the nodes in $ \mathcal{N}(\textbf{s}) $ have the same sensitivity.

\vspace{-5pt}
\begin{definition}[K-norm Mechanism \cite{Geometry-Hardt-STOC10}]
\label{def-knorm}
	Given any function $f(\textbf{s})$: $\textbf{s}\rightarrow \mathbb{R}^d$ and its sensitivity hull $K$, $K$-norm mechanism outputs $\textbf{z}$ withh probability below.
	\begin{myAlignSSS}
	\label{eqn-pdf-K-Norm}
	\Pr(\textbf{z})=\frac{1}{\Gamma(d+1)\textsc{Vol}(K/\epsilon)}exp \left( -\epsilon||\textbf{z}-f(\textbf{s})||_K \right)
	\end{myAlignSSS}
	\vspace{-10pt}
	where $\Gamma(\cdot)$ is Gamma function and $\textsc{Vol}(\cdot)$ denotes volume.
\end{definition}

%


\vspace{-25pt}
\begin{algorithm}[H]
	\scriptsize
	\caption{\small Policy-based  Planar Isotropic Mechanism (P-PIM)}
	\begin{algorithmic}[1]
		\Require{$ \epsilon $,  ${\mathcal{G}}$,  the user's true location $\textbf{s}$.}
		\State{Calculate  $K({\mathcal{G}} )= Conv \big(f(\textbf{s}_i)-f(\textbf{s}_j)\big)  $   for all  neighbors $ \textbf{s}_i, \textbf{s}_j \in\mathcal{N}^P(\textbf{s}) $;}
		\State{$\textbf{z}'=f(\textbf{s})+Y$ where $ Y $ is two-dimension noise drawn by Eq.\eqref{eqn-pdf-K-Norm} with sensitivity hull $ K({\mathcal{G}} )$;}
		\State{\textbf{return} a location $\textbf{z} \in \mathcal{S}$ that is closest to $\textbf{z}'$ on the map.}
	\end{algorithmic}
	\label{algo-p-pim}
\end{algorithm}
\vspace{-15pt}


\vspace{-5pt}
\begin{theorem}
\label{thm2}
P-PIM satisfies $\{\epsilon, \mathcal{G}\}$-location privacy.
\end{theorem}
\vspace{-2pt}

We can prove Theorems \ref{thm1} and \ref{thm2}  using Lemma \ref{lm-graph-dist}.
The sensitivity is scaled with the graph-based distance.
We note that directly using Laplace machanism or PIM can satisfy a fully connected policy graph over locations in the constrained domain as shown in Fig.\ref{figure-two-graphs}(b).

\vspace{-2pt}
\begin{theorem}
	\label{thm-PIM-time}
	Algorithm  \ref{algo-p-pim} has the time complexity
	$ O(|\mathcal{C}| \log(h) + h^2 \log(h)) $ where $ h $ is number of vertices on the polygon of $  Conv(\Delta f ^ \mathcal{G}) $.
\end{theorem}

\vspace{-15pt}

\section{Policy Graph under Constrained Domain}
\label{sec-exposure}
\vspace{-10pt}
In this section, we investigate and prevent  the location exposure of a policy graph under constrained domain in Sec. \ref{sec-node-exposure} and \ref{subsec-detect}, respectively; then we repair the policy graph in Sec. \ref{subsec-repair}.
\vspace{-10pt}
\subsection{Location Exposure}

\label{sec-node-exposure}

\begin{figure}[h]
	\centering
			\vspace{-15pt}
	\includegraphics[width=6.5cm]{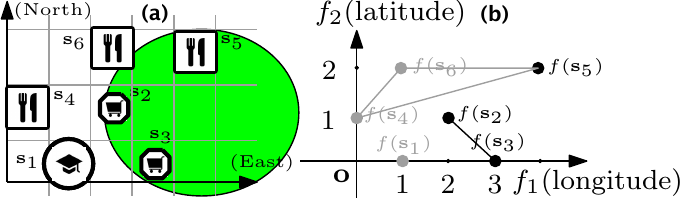}
		\vspace{-5pt}
	\caption{{\small \textbf{(a)} The constrained domain $ \mathcal{C} = \{ \textbf{s}_2, \textbf{s}_3, \textbf{s}_5 \} $; \textbf{(b)} The constrained policy graph. }}
	\label{fig-ex-exposure}
	\vspace{-20pt}
\end{figure}

As shown in Fig.\ref{fig:ex1} (right) and introduced in Section \ref{sec-intro}, a given policy graph may not be viable under adversarial knowledge of constrained domain (Definition \ref{def-constrained-domain}).
We illustrate the potential risks due to the constrained domain shown in Fig.\ref{fig-ex-exposure}.

We first examine the immediate consequences of the constrained domain to the  policy graph by defining the excluded and disconnected nodes.
We then show the disconnected node may lead to \textit{location exposure}.

\vspace{-5pt}
\begin{definition}[Excluded node]
Given a location policy graph $\mathcal{G}=(\mathcal{S},\mathcal{E})$ and a constrained domain $\mathcal{C}\subset  \mathcal{S} $,
if $ s \in  \mathcal{S}$ and $ s \not\in \mathcal{C} $, $\textbf{s}$ is an excluded node.
\end{definition}

\vspace{-10pt}
\begin{definition}[Disconnected node]
Given a  location policy graph $\mathcal{G}=(\mathcal{S},\mathcal{E})$ and a constrained domain $\mathcal{C}\subset  \mathcal{S} $, 
if a node $ \textbf{s} \in \mathcal{C}$, $\mathcal{N}(\textbf{s}) \neq \emptyset  $
 and   { $\mathcal{N}(\textbf{s}) \cap \mathcal{C} = \emptyset$,  we call $ \textbf{s} $} a disconnected node.
\end{definition}

\vspace{-5pt}
Intuitively,  the \textit{excluded} node is outside of the constrained domain $\mathcal{C}$, such as the gray nodes $\{ \textbf{s}_1, \textbf{s}_4, \textbf{s}_6 \}$ in  Fig.\ref{fig-ex-exposure}; whereas the \textit{disconnected} node (e.g., $\textbf{s}_5$ in  Fig.\ref{fig-ex-exposure}) is inside of  $\mathcal{C}$ and has neigbors,  yet all its neighbors are outside of $\mathcal{C}$.

Next, we analyze the feasibility of a location policy graph under a constrained domain.
The first problem is that, by the definition of excluded nodes, it is not possible to achieve indistinguishability between the excluded nodes and any other nodes. 
For example in Fig.\ref{fig-ex-exposure},  the indistinguishability indicated by the gray edges is not feasible because of {\small $\Pr(\mathcal{A}(\textbf{s}_4)=\textbf{z}) = \Pr(\mathcal{A}(\textbf{s}_6)=\textbf{z})=0$} for any $\textbf{z}$ given the adversarial knowledge of {\small $\Pr(\textbf{s}_4) = \Pr(\textbf{s}_6) = 0$}.
Hence, one can only achieve a \textit{constrained policy graph}, such as the one with nodes {\small $\{\textbf{s}_2, \textbf{s}_3, \textbf{s}_5\}$} in Fig.\ref{fig-ex-exposure}(b).

\begin{definition}[Constrained Location Policy Graph]
	\label{def-con-graph}
	A constrained  location policy graph  $\mathcal{G}^\mathcal{C}  $ is a subgraph of the original location policy graph $\mathcal{G}$ under a constrained domain $ \mathcal{C} $ that only includes the edges inside of $\mathcal{C}$. 
	Formally, $\mathcal{G}^\mathcal{C} =(\mathcal{C},\mathcal{E}^\mathcal{C})$ where $  \mathcal{C} \subseteq \mathcal{S}$  and $  \mathcal{E}^\mathcal{C} \subseteq \mathcal{E}$.
\end{definition}

\begin{definition}[Location Exposure under constrained domain]
	Given a  policy graph $\mathcal{G}$, constrained domain $\mathcal{C}$ and an algorithm $\mathcal{A}$ satisfying $(\epsilon, \mathcal{G}^\mathcal{C})$-location privacy, for a disconnected node $\textbf{s}$,
	if $\mathcal{A}$ does not guarantee  $\epsilon$-indistinguish-ability between $\textbf{s}$ and any other nodes in  $\mathcal{C}$, we call  $\textbf{s}$ an isolated node.
	The user suffers location exposure when she is at the location of the isolated node.
\end{definition}

\vspace{-20pt}
\subsection{Detecting Isolated Node}
\label{subsec-detect}

\vspace{-5pt}

An interesting finding is that a disconnected node may not always lead to location exposure, which also depends on the algorithm for PGLP.
Intuitively,  the indistinguishability between a disconnected node and a node in the constrained domain could be guaranteed implicitly. 
We design Algorithm \ref{alg-exposure} to detect the isolated node in a constrained policy graph w.r.t. P-PIM.
It could be extended to any other PGLP mechanism.
For each disconnected node, we check whether it is indistinguishable with other nodes.
The problem is equivalent to checking if there is any node inside the convex body $f(\textbf{s}_{i})+ K({\mathcal{G}^\mathcal{C}})$, which can be solved by the convexity property (if a point $\textbf{s}_{j}$ is inside a convex hull $K$, then $\textbf{s}_{j}$ can be expressed by the vertices of $K$ with coefficients in $[0,1]$) with complexity $O(m^{3})$. 
We design a faster method with complexity $O(m^{2}log(m))$ by exploiting the definition of convex hull: if $\textbf{s}_{j}$ is inside $f(\textbf{s}_{i})+ K({\mathcal{G}^\mathcal{C}})$, then the new convex hull of the new graph by adding edge $\overline{\textbf{s}_{i}\textbf{s}_{j}}$ will be the same as $K({\mathcal{G}^\mathcal{C}})$.
We give an example of \textit{disconnected but not isolated} node in appendix \ref{appx-iso}. 

\vspace{-18pt}
\begin{algorithm}[H]
\scriptsize
	\caption{\small Finding Isolated Node}
	\begin{algorithmic}[1]
		\Require{${\mathcal{G}}$, $\mathcal{C}$, disconnected node $\textbf{s}_i\in \mathcal{C}$.}
		\State{$\Delta f ^\mathcal{G}=\mathop\lor\limits_{\overline{\textbf{s}_j\textbf{s}_k}\in \mathcal{E}^\mathcal{C}}
			\left( f({\textbf{s}_j})-f({\textbf{s}_k}) \right)$;}
		\Comment{We use $ \lor $ to denote Union operator.}
		\State{$K({\mathcal{G}^\mathcal{C}})\gets Conv(\Delta f ^\mathcal{G})$;}
		\ForAll{$\textbf{s}_j\in\mathcal{C},\textbf{s}_j\neq \textbf{s}_i$}
		\If{$Conv(\Delta f ^\mathcal{G}, f(\textbf{s}_j)-f(\textbf{s}_i))==K({\mathcal{G}^\mathcal{C}})$}
		\State{\textbf{return} false}
		\Comment{{\tt \scriptsize not isolated}}
		\EndIf
		\EndFor
		\\\Return{true}
		\Comment{{\tt \scriptsize isolated}}
	\end{algorithmic}
	\label{alg-exposure}
\end{algorithm}

\vspace{-35pt}
\subsection{Repairing Location Policy Graph}
\label{subsec-repair}
\vspace{-5pt}

To prevent location exposure under the constrained domain, we need to make sure that there is no isolated node in a constrained policy graph.
A simple way  is to  modify the  policy graph to ensure the indistinguishability of  the isolated node by adding an edge between it and another node in the constrained domain.
The selection of such a node could depend on the requirement of the application.
Without the loss of generality, a baseline method for repairing the policy graph could be choosing an \textit{arbitrary} node from the constrained domain and adding an edge between it and the isolated node.

A natural question is how can we repair the policy graph with better utility.
Since different ways of adding edges in the policy graph may lead to distinct graph-based sensitivity, which is propositional to the area of the sensitivity hull (i.e., a polygon on the map), the question is equivalent to adding an edge with the \textit{minimum}  area of sensitivity hull (thus the least noise).
We design Algorithm \ref{alg-repair} to find the {minimum} area of the new sensitivity hull, as shown in an example in Fig.\ref{Figure-Gt-example}. 
The  analysis is shown in Appendix \ref{appx-repair}.

We note that both Algorithms \ref{alg-exposure} and \ref{alg-repair} are oblivious to the true location, so they do not consume the privacy budget.
Additionally, the adversary may be able to ``reverse'' the process of graph repair and extract the information about the original location policy graph; however, this does not compromise our privacy notion since the location policy graph is public in our setting.

\begin{figure*}
	\centering
		\vspace{-18pt}
		\includegraphics[width=10cm]{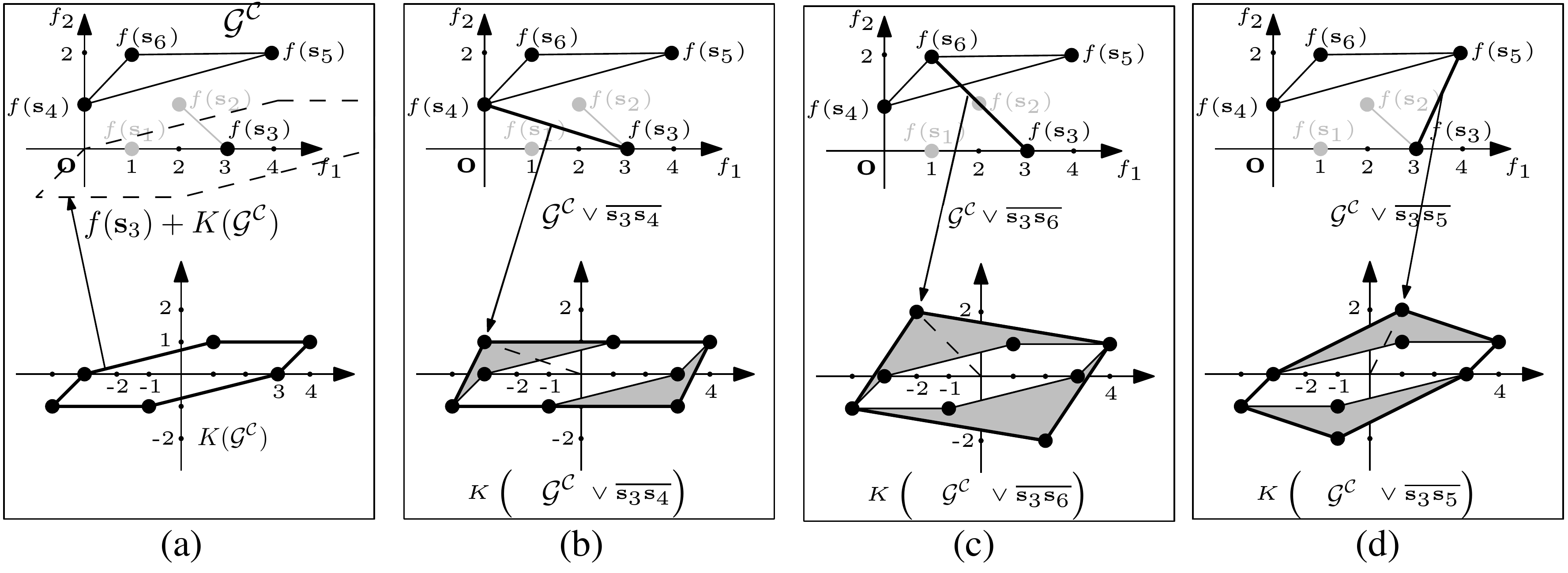}
		\vspace{-10pt}
	\caption{{\small An example of graph repair with high utility.
			(a): if $\mathcal{C}=\{\textbf{s}_3,\textbf{s}_4,\textbf{s}_5,\textbf{s}_6\}$, then $\textbf{s}_3$ is isolated because $f(\textbf{s}_{3})+K(\mathcal{G}^\mathcal{C})$ only contains $\textbf{s}_{3}$; to protect $\textbf{s}_{3}$, we can re-connect $\textbf{s}_{3}$ to one of the valid nodes $\{\textbf{s}_{4},\textbf{s}_{5},\textbf{s}_{6}\}$. 
			(b) shows the new sensitivity hull after adding $f(\textbf{s}_4)-f(\textbf{s}_3)$ to $K(\mathcal{G}^\mathcal{C})$; 
			(c) shows the new sensitivity hull  after adding $f(\textbf{s}_6)-f(\textbf{s}_3)$ to $K(\mathcal{G}^\mathcal{C})$; 
			(d) shows the new sensitivity hull  after adding $f(\textbf{s}_5)-f(\textbf{s}_3)$ to $K(\mathcal{G}^\mathcal{C})$. Because (b) has the smallest {area} of the sensitivity hull, $\textbf{s}_{3}$ should be connected to $\textbf{s}_{4}$.   }}
	\label{Figure-Gt-example}
	\vspace{-15pt}
\end{figure*}

\vspace{-30pt}
\begin{algorithm}[H]
	\scriptsize
	\caption{Graph Repair with High Utility}
	\begin{algorithmic}[1]
		\Require{
			$\mathcal{G}$, $\mathcal{C}$, isolated node $\textbf{s}_{i}$
		}
		\State{$\mathcal{G}^\mathcal{C} \gets \mathcal{G}\land\mathcal{C}$;}
		\State{$K\gets K(\mathcal{G}^\mathcal{C})$;}
		\State{$\textbf{s}_k\gets \emptyset$;}
		\State{$min{Area}\gets \infty$;}
		\ForAll{$\textbf{s}_j\in\mathcal{C}, \textbf{s}_{j}\neq \textbf{s}_{i}$}
		\State{$K\gets K(\mathcal{G}^\mathcal{C}\lor\overline{\textbf{s}_i\textbf{s}_j})$;}
		\Comment{{\tt \scriptsize new sensitivity hull in $O(mlog(m))$}}
		\State{${Area}=\mathop\sum\limits_{i=1,j=i+1}^{i=h}det(\textbf{v}_i,\textbf{v}_j)$ where $\textbf{v}_{h+1}=\textbf{v}_1$;}
		\Comment{{\tt \scriptsize $\Theta(h)$ time}}
		\If{${Area}<min{Area}$}
		\State{$\textbf{s}_k\gets\textbf{s}_j$;}
		\State{$min{Area}={Area}$;}
		\Comment{{\tt \scriptsize find minimum area}}
		\EndIf
		\EndFor
		\State{$\mathcal{G}^\mathcal{C}\gets\mathcal{G}^\mathcal{C} \lor\overline{\textbf{s}_i\textbf{s}_k}$}
		\Comment{\tt \scriptsize add edge $ \overline{\textbf{s}_i\textbf{s}_k}  $ to the graph}
		\\\Return{repaired policy graph $\mathcal{G}^\mathcal{C}$;}
	\end{algorithmic}
	\label{alg-repair}
\end{algorithm}

\vspace{-25pt}
\section{Location Trace Release with PGLP}
\label{sec-release}
\vspace{-10pt}

\subsection{Location Release via Hidden Markov Model}
\vspace{-5pt}

\begin{figure}[H]
	\vspace{-20pt}
	\centering
	\begin{subfigure}{0.6\textwidth}
		\includegraphics[width=7cm]{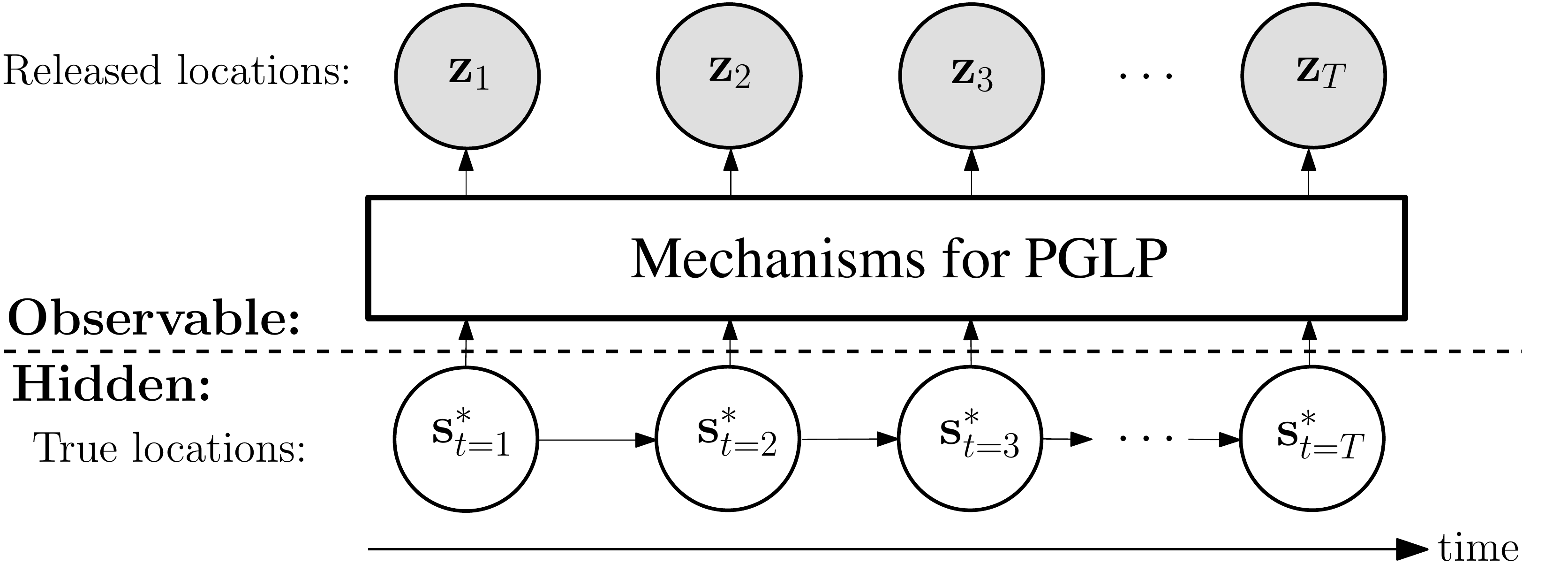}
	\end{subfigure}
\vspace{-5pt}
	\caption{{\small Private location trace release via HMM.}}
	\vspace{-25pt}
	\label{Figure-HMM}
\end{figure}

A remaining question for continuously releasing private location with PGLP is how to calculate the adversarial knowledge of constrained domain $ \mathcal{C}_t $  at each time $ t $.
According to our adversary model described in Sec. \ref{sec-prob-state}, the attacker knows the user's mobility pattern modeled by the Markov chain and the initial probability distribution of the user's location.
The attacker also knows the released mechanisms for PGLP. 
Hence, the problem of calculating the possible location domain  (i.e., locations that {\small $\Pr(\textbf{s}_t^* )>0$}) can be modeled as an inference problem in Hidden Markov Model (HMM)  in Figure \ref{Figure-HMM}: the attacker attempts to infer the  probability distribution of the true location $\textbf{s}_t^* $, given the PGLP mechanism, the Markov model of $\textbf{s}_t^* $, and the observation of $ \textbf{z}_1, \cdots, \textbf{z}_t$ at the current time $ t $.
The constrained domain at each time is derived as the locations in the probability distribution of $\textbf{s}_t^* $ with non-zero probability.

We elaborate the calculation of the probability distribution of $\textbf{s}_t^* $ as follows.
The probability $\Pr(\textbf{z}_{t}|\textbf{s}^*_{t})$ denotes the distribution of the released location $\textbf{z}_t$ where $\textbf{s}^*_{t}$ is the true location at any timestamp $t$.
At timestamp $t$, we use $\textbf{p}_t^-$ and $\textbf{p}_t^+$  to denote the prior and posterior probabilities of an adversary about current state before and after observing $\textbf{z}_t$ respectively.
The prior probability can be derived by the (posterior) probability at  previous timestamp $t-1$
and the Markov transition matrix as
$
\textbf{p}_{t}^-=\textbf{p}_{t-1}^+\textbf{M}
$.
The posterior probability can be computed using Bayesian inference as follows. For each state $\textbf{s}_i$:
\begin{myAlignSSS}
\textbf{p}_{t}^+[i]=\Pr(\textbf{s}_{t}^*=\textbf{s}_i|\textbf{z}_{t})
=\frac{\Pr(\textbf{z}_{t}|\textbf{s}_{t}^*=\textbf{s}_i)\textbf{p}_{t}^-[i]}{\sum_{j}Pr(\textbf{z}_{t}|\textbf{s}_{t}^*=\textbf{s}_j)\textbf{p}_{t}^-[j]}
\label{eqn-posterior}
\end{myAlignSSS}

\vspace{-10pt}

Algorithm \ref{alg-framework} shows the location trace release algorithm.
At each timestamp $t$, we compute the constrained domain (Line 2).
For all disconnected nodes under the constrained domain, we check if they are isolated by Algorithm \ref{alg-exposure}. 
If so, we derive a minimum protectable graph $\mathcal{G}_t$ by Algorithm \ref{alg-repair}.
Next, we use the proposed PGLP mechanisms (i.e., P-LM or P-PIM)  to release a perturbed location $\textbf{z}_t$.
Then the released $\textbf{z}_t$ will also be used to update the posterior probability $\textbf{p}_t^+$ (in the equation below) by Equation (\ref{eqn-posterior}), which subsequently will be used to compute the prior probability for the next time $t+1$.
We note that, only Line 9 (invoking PGLP mechanisms) uses the true location $ \textbf{s}_t^* $.
Algorithms  \ref{alg-exposure} and \ref{alg-repair} are independent of the true location, so they do not consume the privacy budget.

\vspace{-20pt}
\begin{algorithm}[H]
	\caption{{\small Location Trace Release Mechanism for PGLP}}
	\scriptsize
	\begin{algorithmic}[1]
		\Require{
			$\epsilon$, $\mathcal{G}$, $\textbf{M}$, $\textbf{p}_{t-1}^+$, $\textbf{s}_t^*$
		}
		\State{$\textbf{p}_{t}^-\gets\textbf{p}_{t-1}^+\textbf{M}$;}
		\Comment{{\tt \scriptsize Markov transition}}
		\State{$\mathcal{C}_t\gets\{\textbf{s}_i|\textbf{p}_t^-[i]>0\}$;}
		\Comment{{\tt \scriptsize constraint}}
		\State{$\mathcal{G}_{t}^\mathcal{C}\gets \mathcal{G}\land\mathcal{C}_{t}$;}
		\Comment{{\tt Definition \ref{def-con-graph}}}
		\ForAll{disconnected node $\textbf{s}_{i}$ in $\mathcal{G}_{t}^\mathcal{C}$}
		\If{$\textbf{s}_{i}$ is isolated }
		\Comment{{\tt \scriptsize isolated node detection by Algorithm \ref{alg-exposure}}}
		\State{$\mathcal{G}_t^\mathcal{C} \gets$ \Call{Algorithm \ref{alg-repair}}{$\mathcal{G}_{t}^\mathcal{C}$, $\mathcal{C}_t$, $\textbf{s}_{i}$};}
		\Comment{{\tt \scriptsize repair graph $\mathcal{G}_t$ by Algorithm \ref{alg-repair}}}
		\EndIf
		\EndFor
		\State{mechanisms for PGLP with parameters $ \epsilon$, $\textbf{s}_t^*$, ${\mathcal{G}}_t$;}
		\Comment{{\tt Algorithms \ref{algo-p-lm} or \ref{algo-p-pim}}}
		\label{line-releasing}
		\State{Derive $\textbf{p}_t^+$ by Equation (\ref{eqn-posterior});}
		\Comment{{\tt \scriptsize inference go to next timestamp}}
		\label{line-bayesian}
		\\\Return{\Call{Algorithm \ref{alg-framework}}{$\epsilon$, $\mathcal{G}_t^\mathcal{C}$, $\textbf{M}$, $\textbf{p}_{t}^+$, $\textbf{s}_{t+1}^*$}};
	\end{algorithmic}
	\label{alg-framework}
\end{algorithm}

\vspace{-20pt}

\begin{theorem}[Complexity]
Algorithm \ref{alg-framework} has complexity $O(dm^{2}log(m))$ where $d$ is the number of disconnected nodes and $m$ is the number of nodes in $\mathcal{G}_t^\mathcal{C}$. 
\end{theorem}

\subsection{Privacy Composition}
\label{sec-priv-composition}
We analyze the  composition of privacy  for multiple location releases under PGLP.
In Definition \ref{def-PGLP}, we define $ \{\epsilon, \mathcal{G}\}$-location privacy for single location release, where $ \epsilon $ can be considered the privacy leakage w.r.t. the privacy policy $ \mathcal{G} $.
A natural question is what would be the privacy leakage of multiple releases at a single timestamp (i.e., for the same true location) or at multiple timestamps (i.e., for a trajectory).
In either case, the privacy guarantee (or the upper bound of privacy leakage) in multiple releases depends on the achievable location  policy graphs.
Hence,  the key is to study the composition of  the policy graphs in multiple releases.  
Let {\small $ \mathcal{A}_1, \cdots, \mathcal{A}_T$} be $ T $ independent random algorithms that 
takes true locations {\small $ \textbf{s}_1^*, \cdots, \textbf{s}_T^*$ } as inputs (note that it is possible {\small $ \textbf{s}_1^* = \cdots = \textbf{s}_T^*$})  and outputs $ \textbf{z}_1, \cdots, \textbf{z}_T $, respectively.
When the viable policy graphs are the same at each release, we have Lemma \ref{thm-comp1} as below.

\begin{lemma}
	\label{thm-comp1}
If all $ \mathcal{A}_1, \cdots, \mathcal{A}_T$ satisfy $ (\epsilon, \mathcal{G} )$-location privacy,
the combination of  \{$ \mathcal{A}_1, \cdots, \mathcal{A}_T$\} satisfies $ (T\epsilon, \mathcal{G} )$-location privacy.
\end{lemma}

 As shown in Sec. \ref{sec-exposure}, the feasibility of  achieving a policy graph  is affected by the constrained domain, which may change along with the released locations.
 We denote  {\small $\mathcal{G}_1, \cdots, \mathcal{G}_T $} as viable policy graphs at each release (for single location or for a trajectory), which could be obtained by algorithms in Sec. \ref{subsec-detect} and Sec. \ref{subsec-repair}.
 We give a more general composition theorem for PGLP below.

\begin{theorem}
	\label{thm-comp2}
If {\small  $ \mathcal{A}_1, \cdots, \mathcal{A}_T$} satisfy  {\small $ (\epsilon_1, \mathcal{G}_1 ), \cdots, (\epsilon_T, \mathcal{G}_T)$}, -location privacy, respectively,
the combination of  {\small $ \{ \mathcal{A}_1, \cdots, \mathcal{A}_T \}$ } satisfies  {\small $ \big(\sum_{i=1}^{T}\epsilon_i,  \mathcal{G}_1 \wedge  \cdots \wedge\mathcal{G}_T \big)$}-location privacy, where $ \wedge   $ denotes the intersection between the edges of policy graphs.
\end{theorem}

The above theorem provides a method to reason about the overall privacy in continuous releases using PGLP.
 We note that the privacy composition does not depend on the adversarial knowledge of Markov model, but replies on the soundness of the policy graph and PGLP mechanisms at each $ t $.
However, the resulting $ \mathcal{G}_1 \wedge  \cdots \wedge\mathcal{G}_T $ may not be the original policy graph.
It is an interesting future work to study how to ensure a given policy graph across the timeline.


\section{Experiments}
\subsection{Experimental Setting}
We implement the algorithms use Python 3.7. 
The code is available in github\footnote{\url{https://github.com/emory-aims/pglp}.}.
We run the algorithms on a machine  with Intel core i7 6770k CPU and  64 GB of memory  running Ubuntu 15.10 OS.
	
\noindent{\bf Datasets.} 
We evaluate the algorithms on three real-world datasets with similar configurations in \cite{xiao_ccs15} for comparison purpose.
The Markov models were learned from the raw data.
For each dataset, we randomly choose $20$ users' location trace with $100$ timestamps for testing.
\begin{itemize}
	\item
	Geolife dataset \cite{GeoLife-Zheng-10} contains tuples with attributes of user ID, latitude, longitude and timestamp.
	We extracted all the trajectories within the Fourth Ring of Beijing to learn the Markov model, with the map partitioned into cells of  {\small $0.34\times 0.34\ {km}^2$}.
	\item
	Gowalla dataset \cite{KDD-Gowalla-2011} contains $6,442,890$ check-in locations of $196,586$ users over 20 months. 
	We extracted all the check-ins in Los Angeles to train the Markov model, with the map partitioned into cells of {\small $0.37\times 0.37\ {km}^2$}.
	\item
	Peopleflow dataset\footnote{\url{http://pflow.csis.u-tokyo.ac.jp/}}  includes $102,468$ locations of $11,406$ users with semantic labels of POI in Tokyo. 
	We partitioned the map into cells of {\small $0.27\times 0.27\ {km}^2$}.
\end{itemize}

\vspace{2mm}\noindent{\bf Policy Graphs.}
We evaluate two types of location privacy policy graphs for different  applications as introduced in Section \ref{sec-intro}.
One  is for the policy of ``\textit{allowing the app to access  a user's location in which area but ensuring indistinguishability among locations in each area}'', represented by $ G_{k9}, G_{k16}, G_{k25} $ below.
The other is for the policy of ``\textit{allowing the app to access  the semantic label (e.g., a restaurant or a shop) of a user's location but ensuring indistinguishability among locations with the same category}'', represented by $ G_{poi} $  below.
\begin{itemize}
	\item
	$G_{k9}$ is a policy graph that all locations in each {\small $ 3\times 3 $} region (i.e.,  9  grid cells using \textit{grid coordinates}) are fully connected with each other.  
	Similarly, we have	{\small $G_{k16}$} and 	{\small $G_{k25}$} for region size {\small $ 4\times 4 $ } and {\small $ 5\times 5 $}, respectively.
	\item
	$G_{poi}$: all locations with both the same category and the same $ 6 \times 6 $ region are fully connected. 
    We test the category of restaurant in Peopleflow dataset.
\end{itemize}

\noindent{\bf Utility Metrics.} We evaluate three types of utility (error)  for different applications.
We run the mechanisms 200 times and average the results.
Note that  the lower value of the following metrics, the better utility.
\vspace{-6pt}
\begin{itemize}
	\item
	The general utility was measured by Euclidean distance (km), i.e., $ E_{eu} $, between the released location and the true location as defined in Sec. \ref{sec-prob-state}.
	\item
	The utility for weather apps or road traffic monitoring, i.e., ``whether the released location is in the same region with the true location''. 
	We measure it by {\small $ E_r = ||{R}(\textbf{s}^*), {R}(\textbf{z})||_0$} where {\small $ {R}(\cdot) $} is a region query that  returns the index of the region. Here we define the region size as $ 5 \times 5 $ grid cells.
	\item
	The utility for POI mining or crowd monitoring during the pandemic, i.e., ``whether the released location is the same category with the true location''.
	We measure it by {\small $ E_{poi} = ||{C}(\textbf{s}^*), {C}(\textbf{z})||_0$} where {\small $ {C}(\cdot) $} returns the category of the corresponding location.  We evaluated the location category of ``restaurant''.
\end{itemize}


\subsection{Results and Analysis}

\begin{figure}[t]
	\centering
	\begin{minipage}{0.3\hsize}
		\centering
		\includegraphics[width=\hsize]{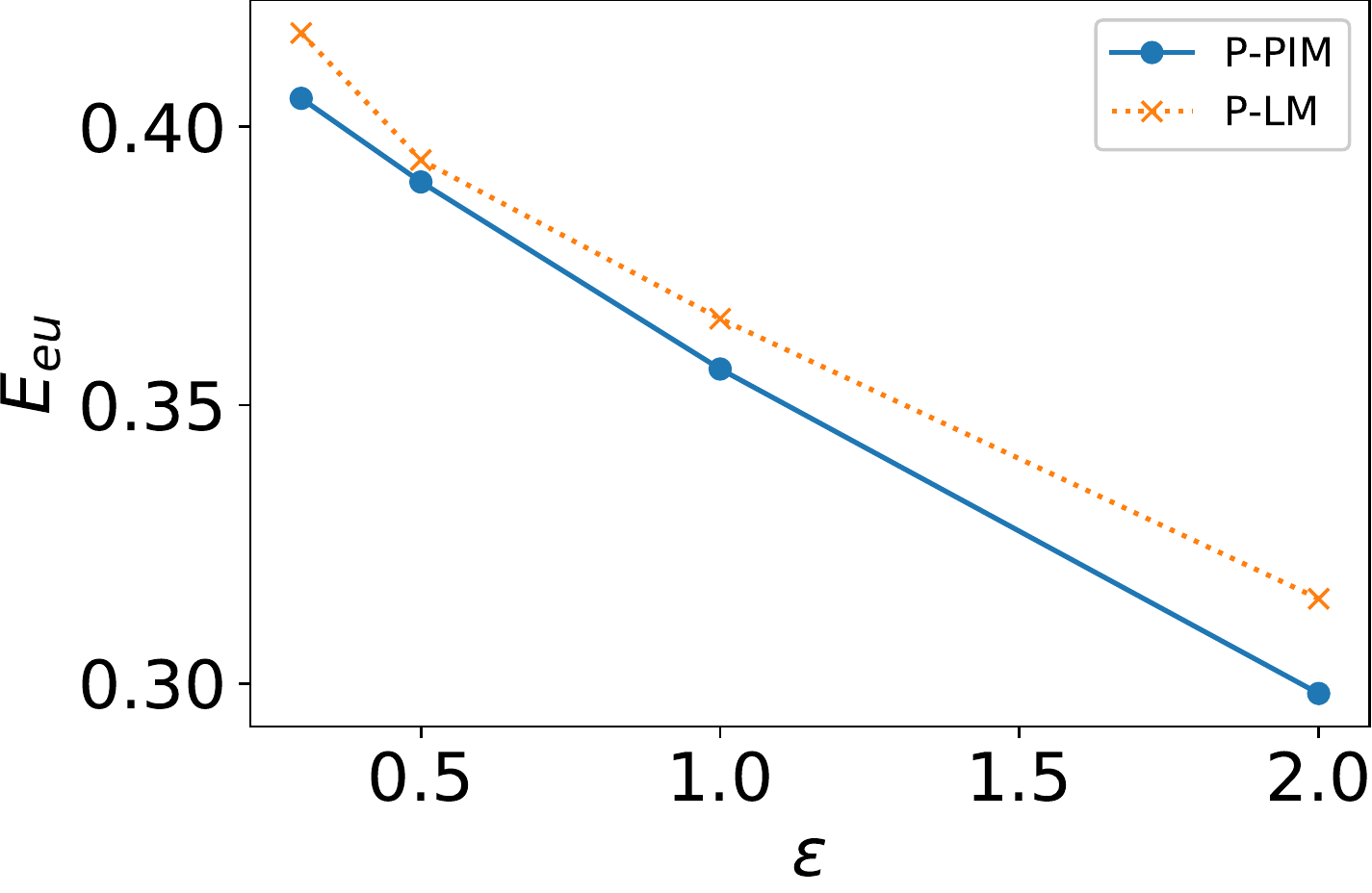}
	\end{minipage}
	\begin{minipage}{0.3\hsize}
		\includegraphics[width=\hsize]{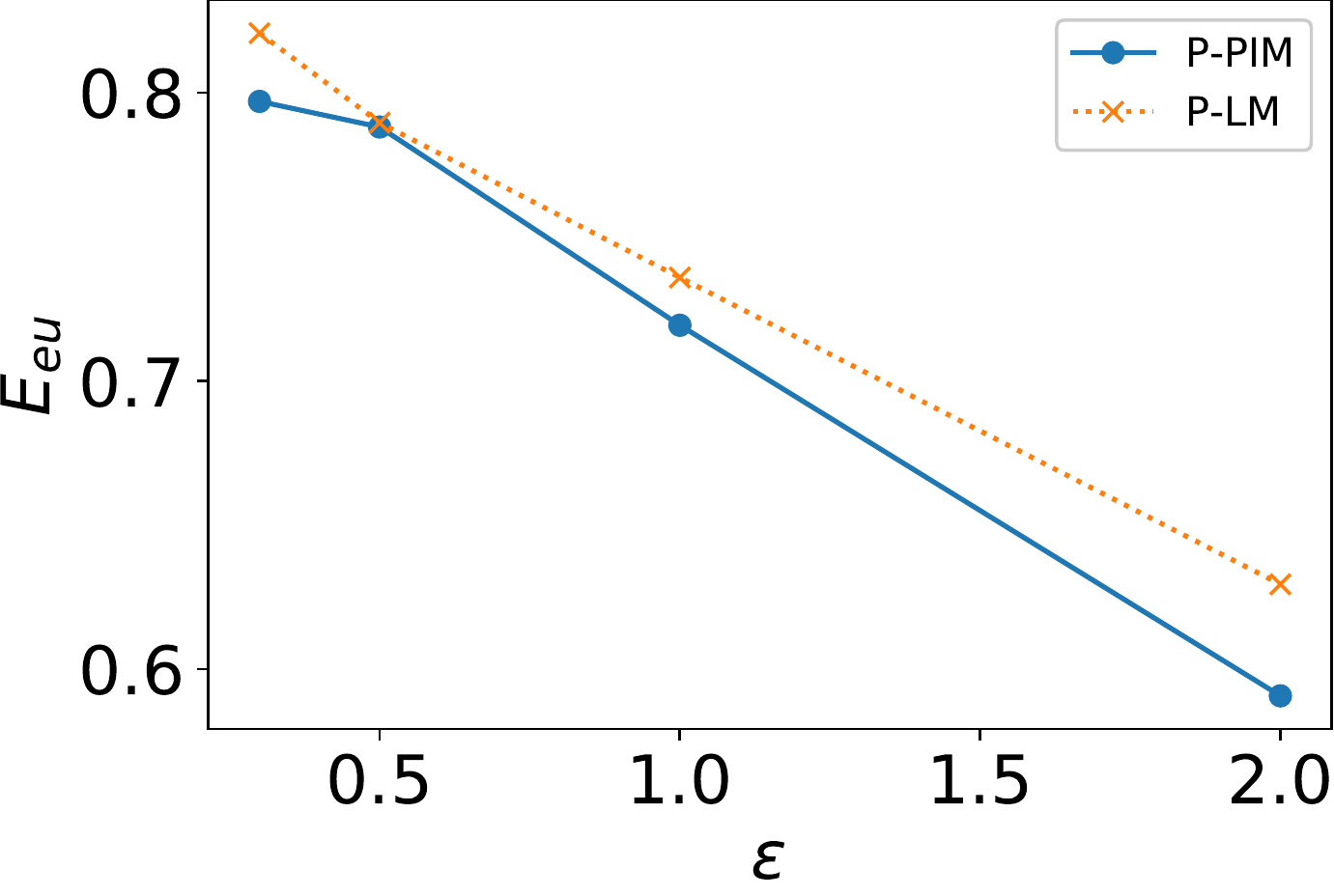}
	\end{minipage}
	\begin{minipage}{0.3\hsize}
		\includegraphics[width=\hsize]{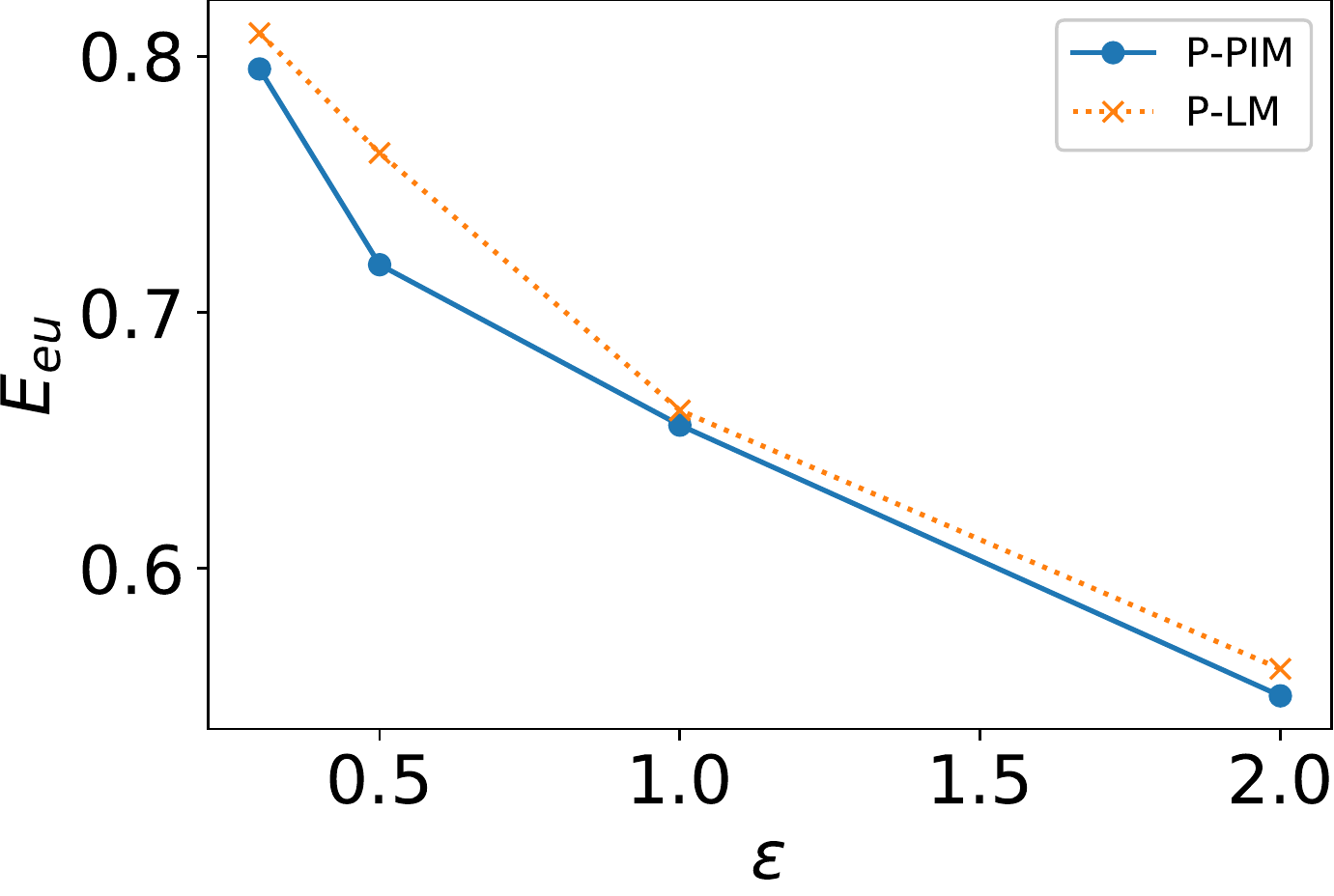}
	\end{minipage}
	\vspace{-10pt}
	\caption{\label{fig:p-lm_vs_p_pim}{\small Utility of P-LM vs. P-PIM with respect to $G_{k9}$, $G_{k16}$ and $G_{poi}$.}}
		\vspace{-5pt}
\end{figure}

\noindent \textbf{P-LM vs. P-PIM}.
Fig.\ref{fig:p-lm_vs_p_pim} compares the utility of two proposed mechanisms P-LM and P-PIM for PGLP under the policy graphs $ G_{k9} $, $ G_{k16} $ , $ G_{k25} $ and $ G_{poi} $ on Peopleflow dataset.
The utility of P-PIM outperforms  P-LM for different policy graphs and different $ \epsilon $ since the sensitivity hull could achieve lower error bound.

\begin{figure}[t]
		\centering
	\begin{minipage}{0.3\hsize}
		\centering
		\includegraphics[width=\hsize]{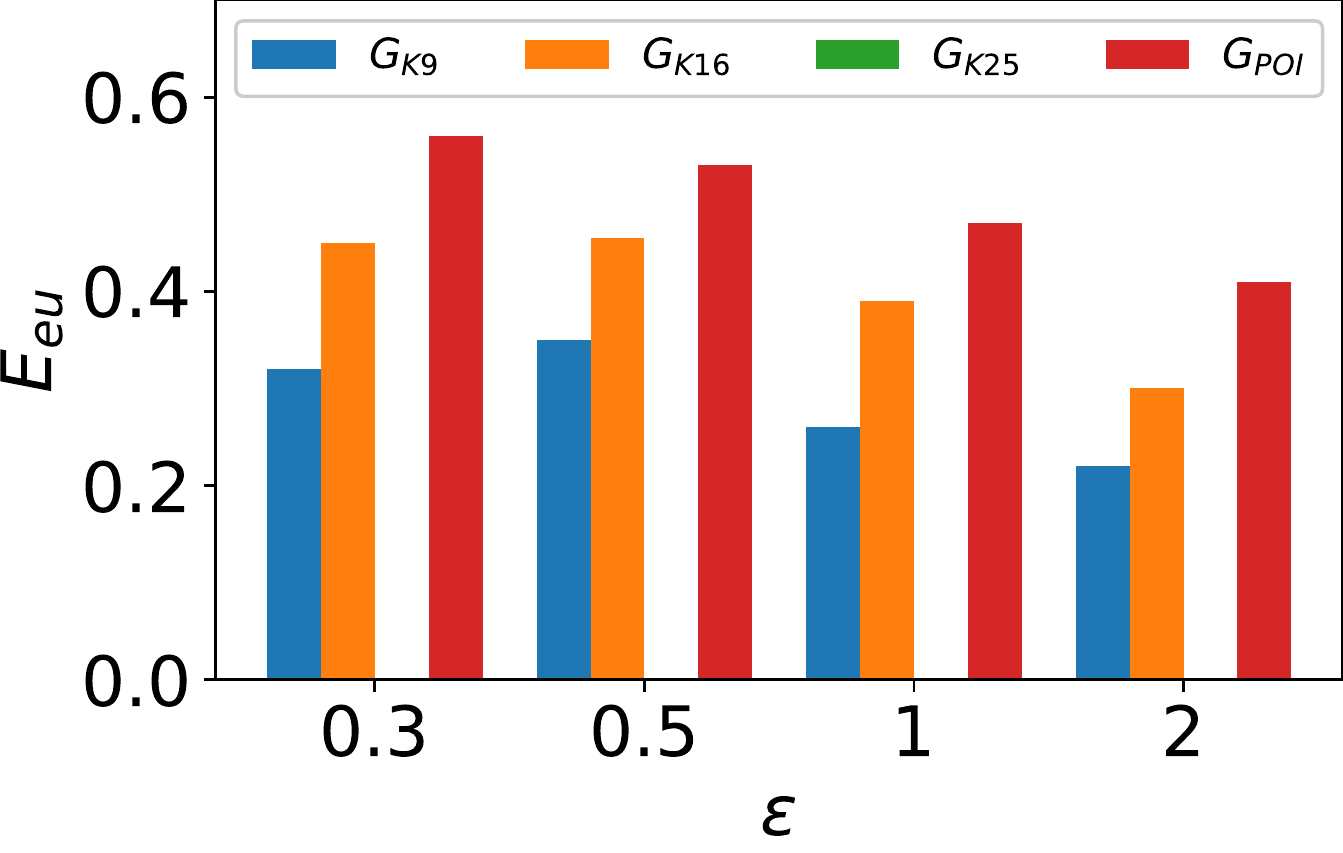}
	\end{minipage}
	\begin{minipage}{0.3\hsize}
		\includegraphics[width=\hsize]{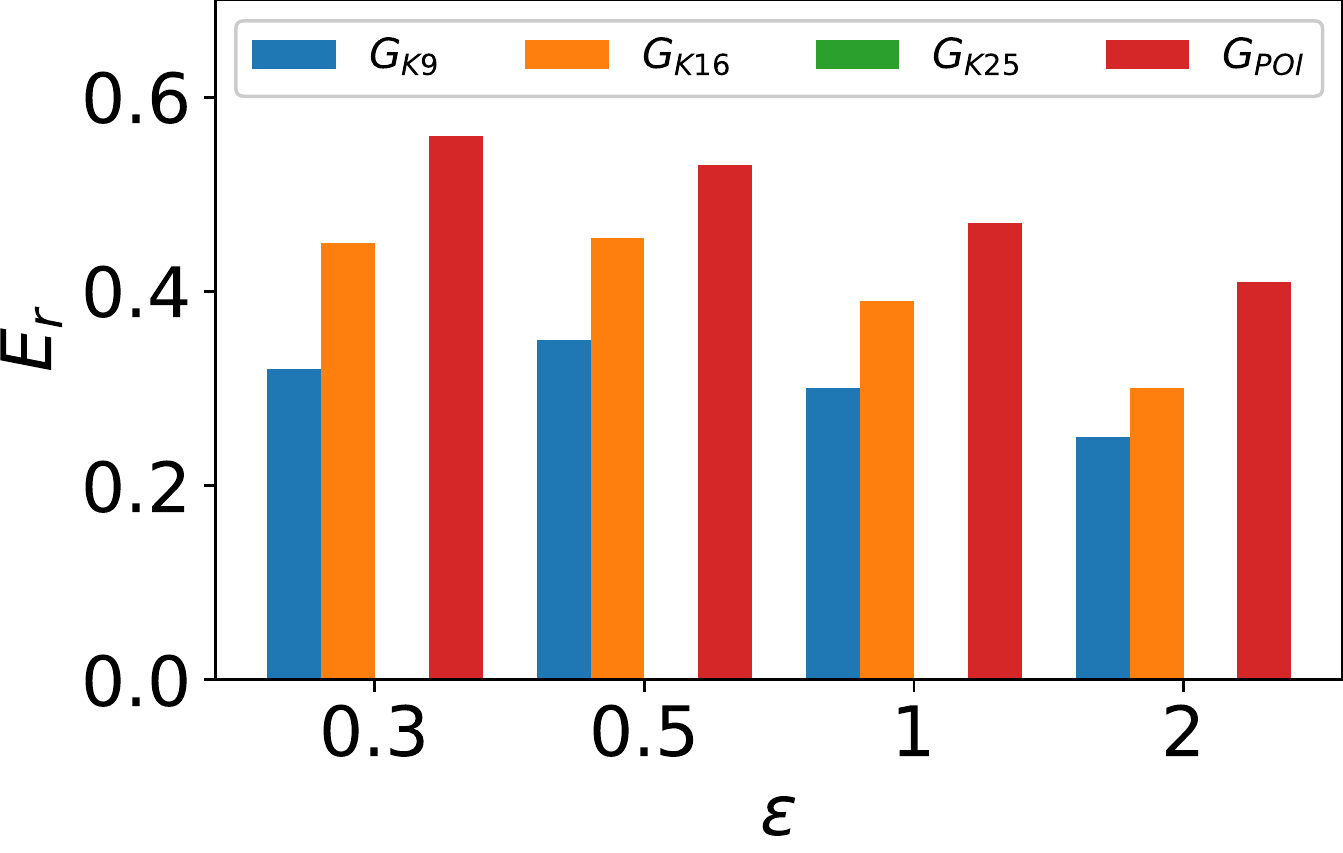}
	\end{minipage}
	\begin{minipage}{0.3\hsize}
		\includegraphics[width=\hsize]{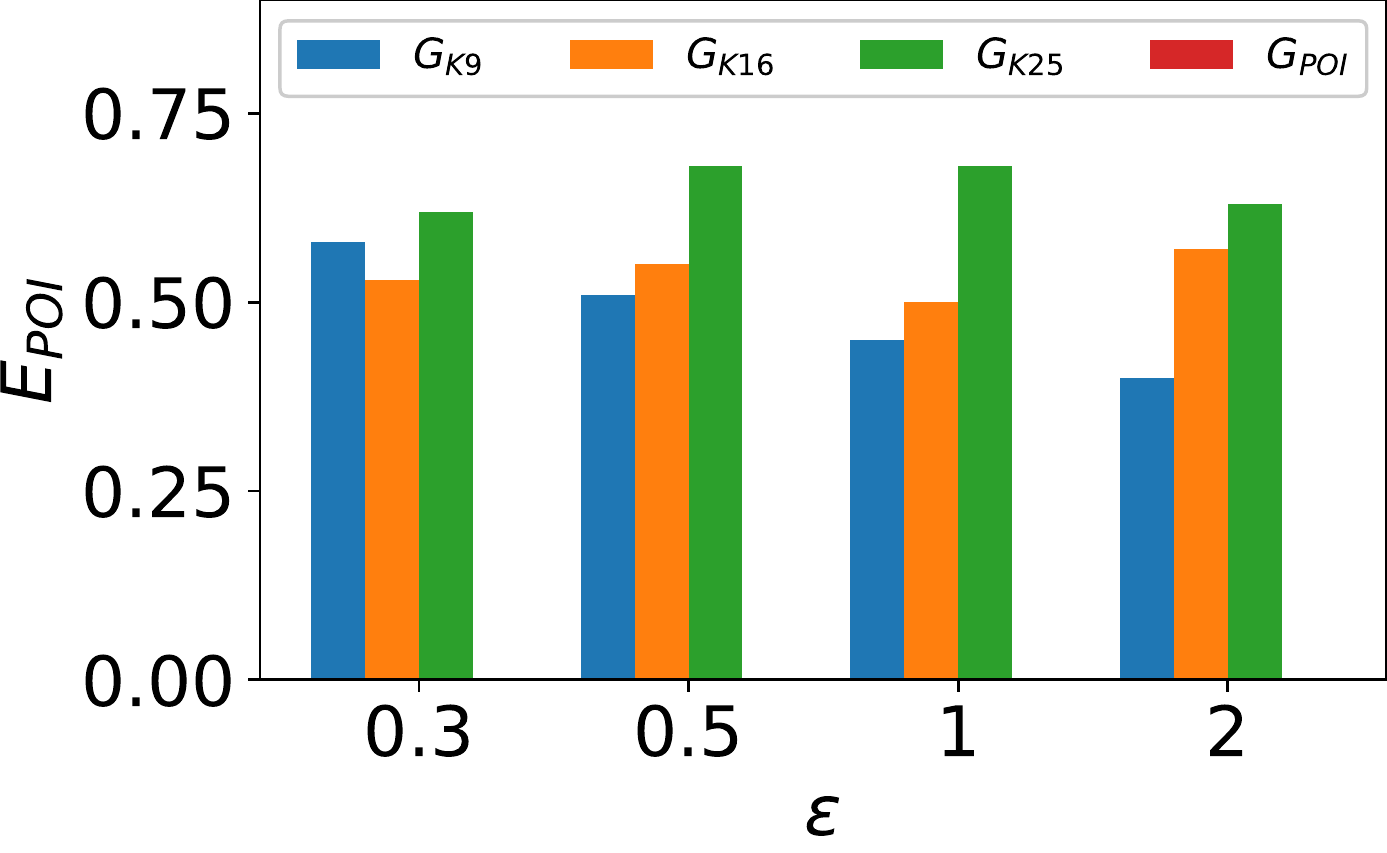}
	\end{minipage}
	\vspace{-10pt}
	\caption{\label{fig:util_on_graphs}{\small Utility of different policy graphs.}}
	\vspace{-15pt}
\end{figure}

\noindent \textbf{Utility Gain by Tuning Policy Graphs.} 
Fig.\ref{fig:util_on_graphs} demonstrates that the utility of different applications can be boosted with appropriate policy graphs.
We evaluate the three types of utility metrics  using different policy graphs on Peopleflow dataset.
Fig.\ref{fig:util_on_graphs} shows that, for utility metrics $ E_{eu} $, $ E_{r} $  and $ E_{poi} $, the policy graphs with the best utility are $ G_{k9} $, $ G_{k25} $   and $ G_{poi} $, respectively. 
$ G_{k9} $ has smallest $ E_{eu} $ because of the least sensitivity.
When the query is $ 5 \times 5 $ region query, $ G_{k25} $  has the full usability ($  E_{r}  $=0).
When the query is POI query like the one mentioned above, $ G_{poi} $  leads to full utility ($  E_{r}  $=0) since  $ G_{poi} $ allows to disclose the semantic category of the true location while maintaining the indistinguishability among the set of locations with the same category.
Note that   $ E_{poi}  $ is decreasing  with larger $\epsilon$ for policy graph $  G_{9} $  because the perturbed location has a higher probability to be the true location; while this effect is diminished in larger policy graphs such as $ G_{16} $ or $  G_{25} $ due to their larger sensitivities.
We conclude that location policy graphs can be tailored flexibly for better utility-privacy trade-off.

\vspace{-15pt}

\begin{figure}[h]
		\centering
	\begin{minipage}{0.3\hsize}
		\centering
		\includegraphics[width=\hsize]{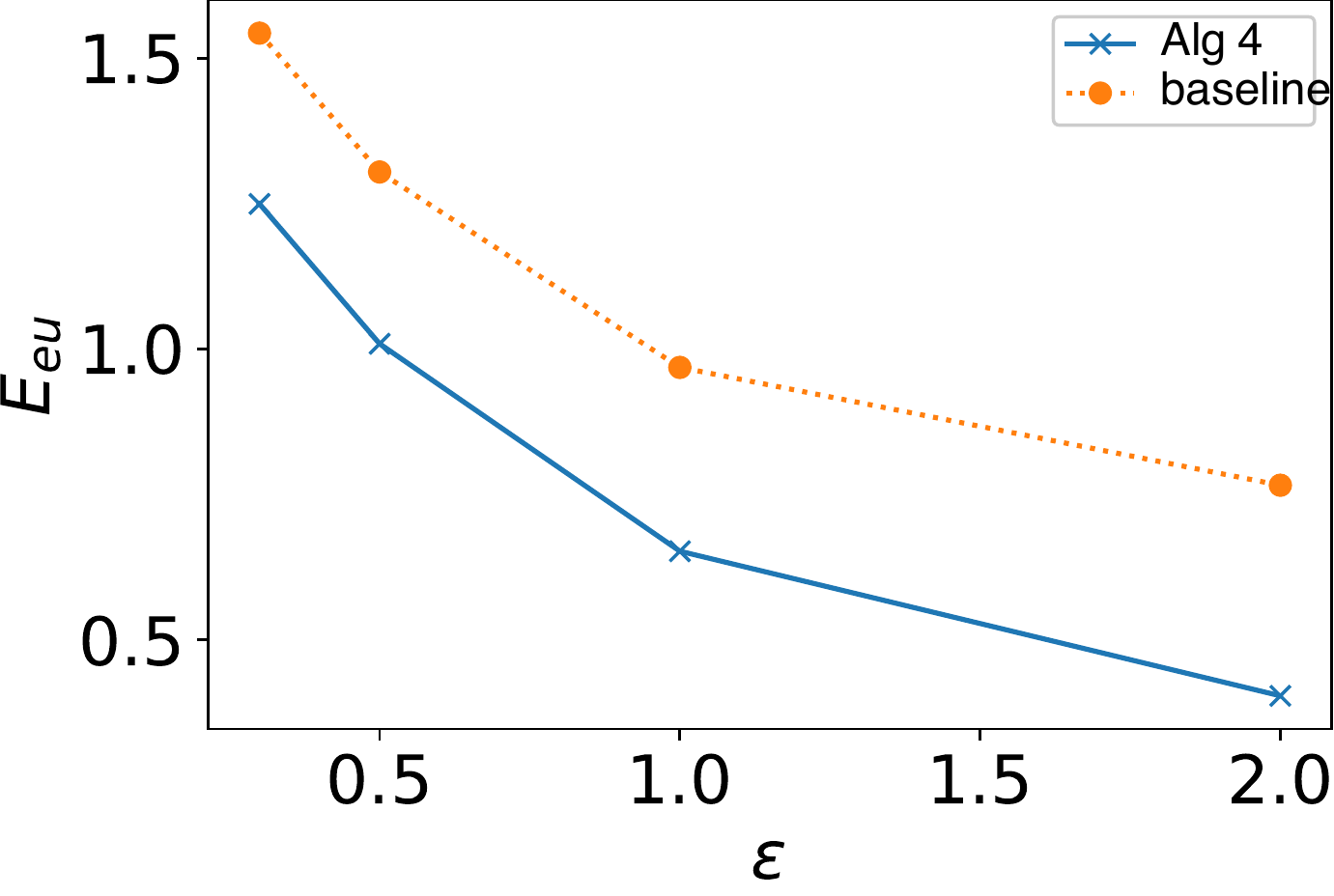}
	\end{minipage}
	\begin{minipage}{0.3\hsize}
		\includegraphics[width=\hsize]{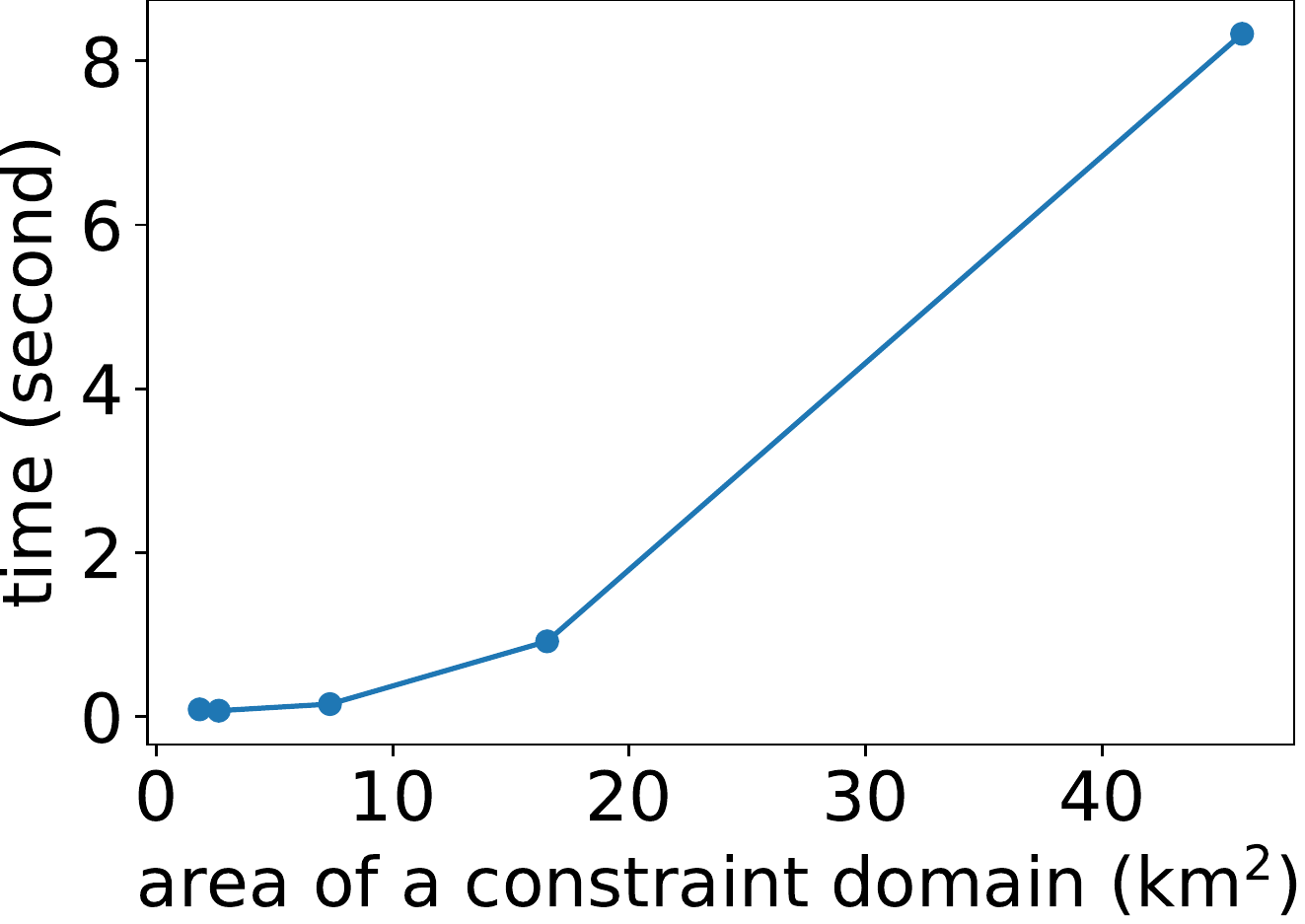}
	\end{minipage}
	\begin{minipage}{0.3\hsize}
		\includegraphics[width=\hsize]{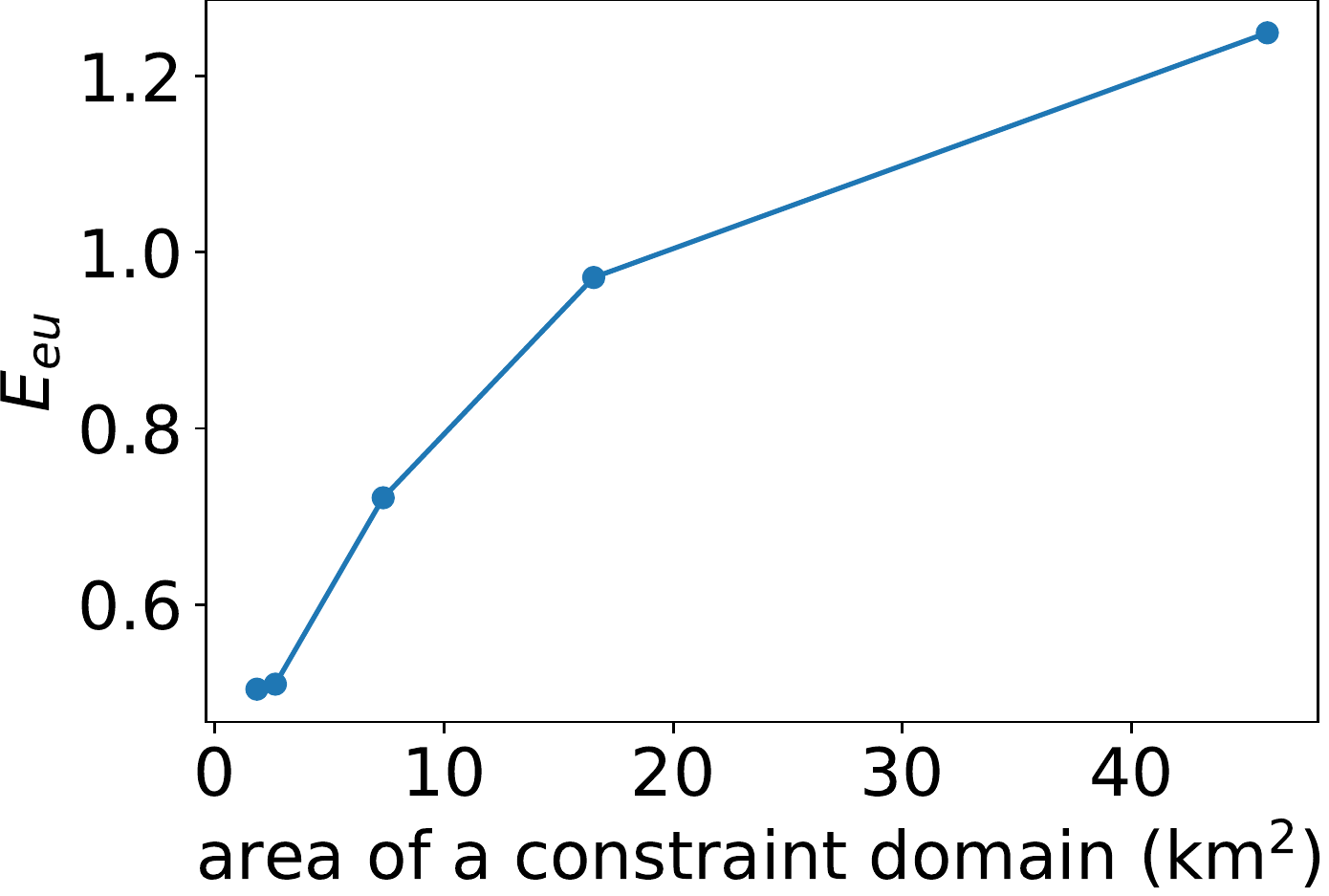}
	\end{minipage}
	\vspace{-5pt}
	\caption{\label{fig:repair_graph_alg5_vs_base}{\small Evaluation of Graph Repair.}}
	\vspace{-20pt}
\end{figure}

\noindent \textbf{Evaluation of Graph Repair.}
Fig. \ref{fig:repair_graph_alg5_vs_base} shows the results of graph repair algorithms.
We compare the proposed Algorithm \ref{alg-repair} with a baseline method that repairs the problematic policy graph by adding an edge between the isolated node with its nearest node in the constrained domain.
It shows that the utility measured by $ E_{eu} $ of Algorithm \ref{alg-repair}  is always better than the baseline but at the cost of higher runtime.
Notably, the utility is decreasing (i.e., higher $ E_{eu} $ ) with larger constrained domains because of larger policy graph (thus higher sensitivity); a larger constrained domain also incurs higher runtime because more isolated nodes need to be processed.

\begin{figure}[t]
	\centering
	\begin{minipage}{0.35\hsize}
		\centering
		\includegraphics[width=\hsize]{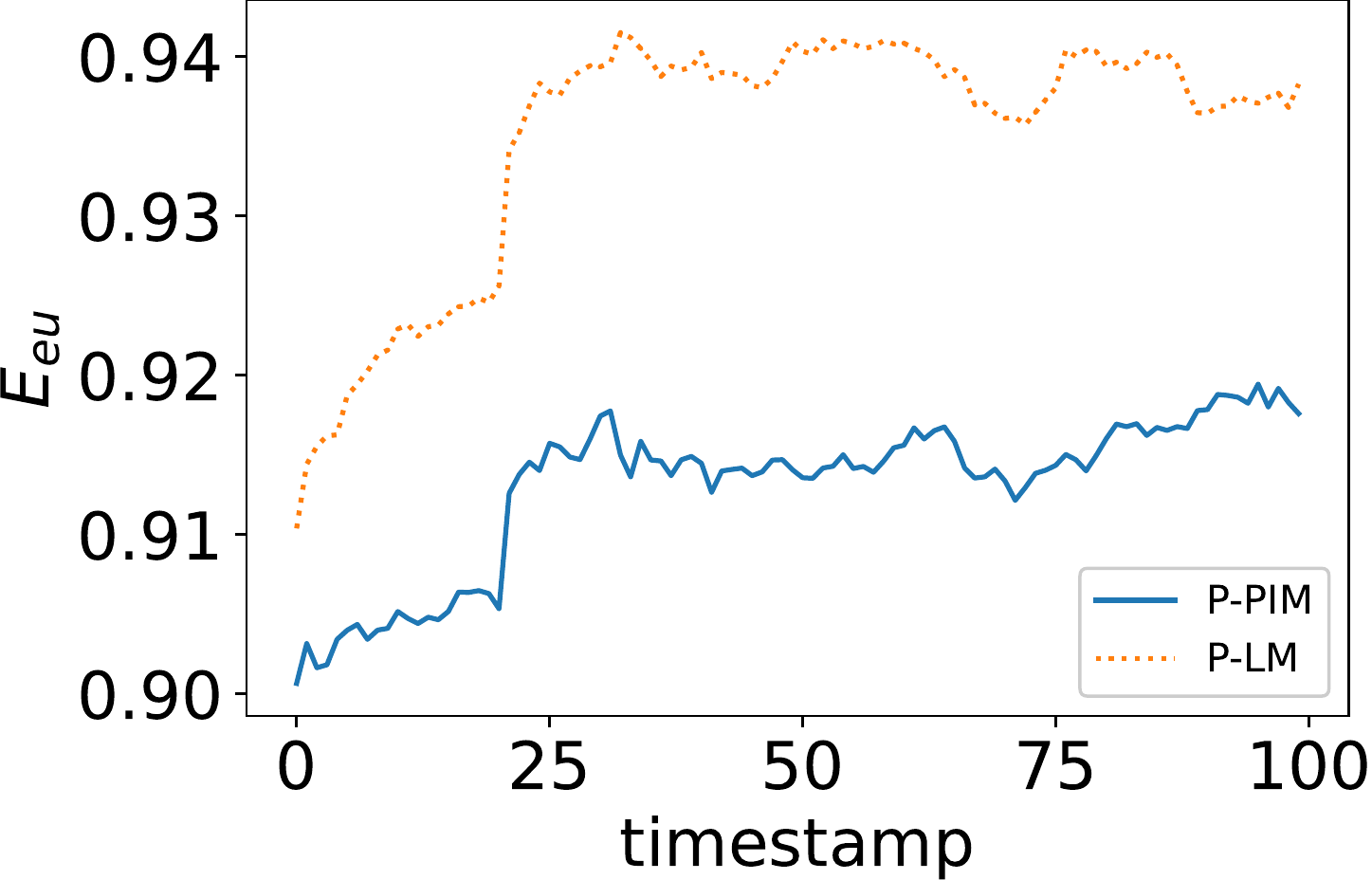}
	\end{minipage}
	\begin{minipage}{0.35\hsize}
		\includegraphics[width=\hsize]{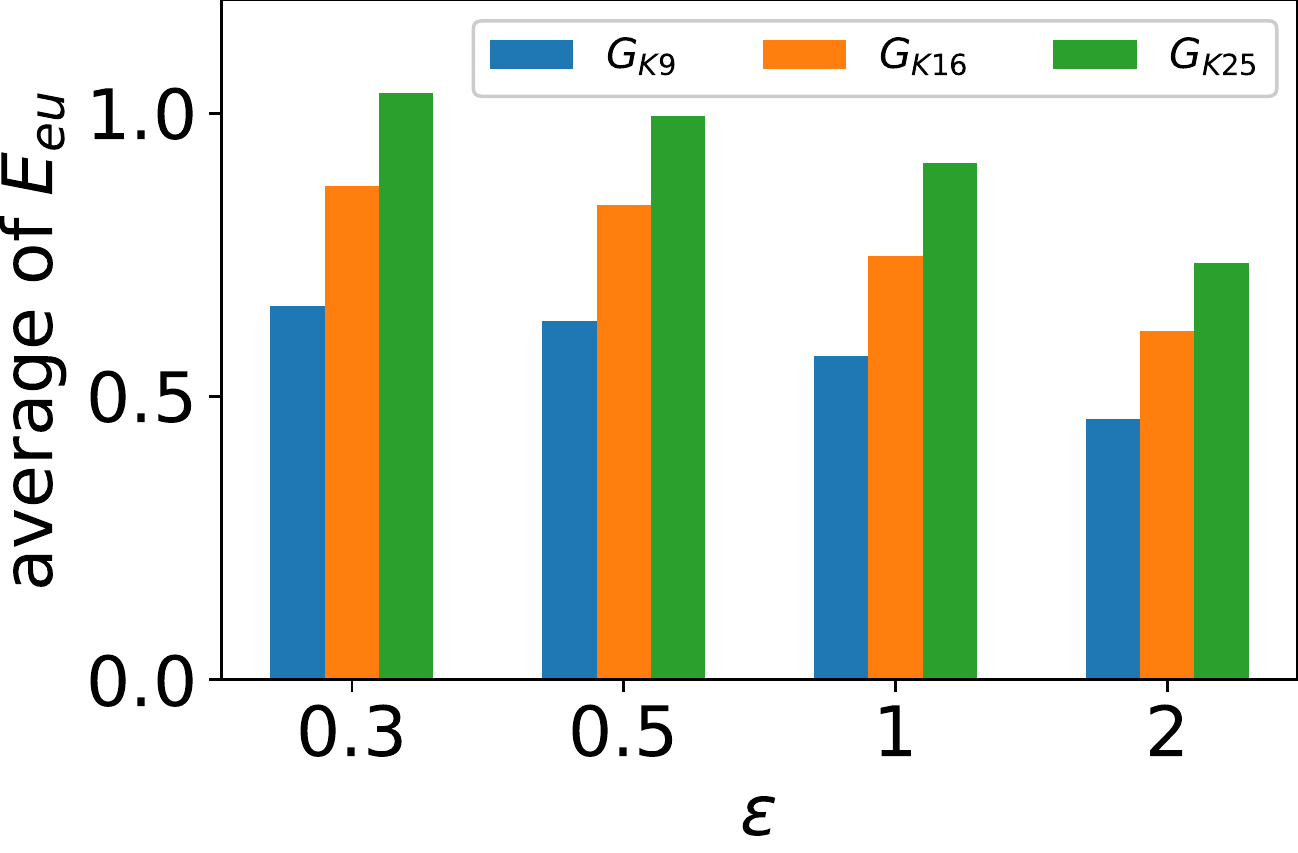}
	\end{minipage}
	\vspace{-10pt}
	\caption{\label{fig-traj1}{\small Utility of Private Trajectory Release with  P-LM and P-PIM.}}
	\vspace{-10pt}
\end{figure}

\begin{figure}[t]
	\centering
	\begin{minipage}{0.35\hsize}
		\includegraphics[width=\hsize]{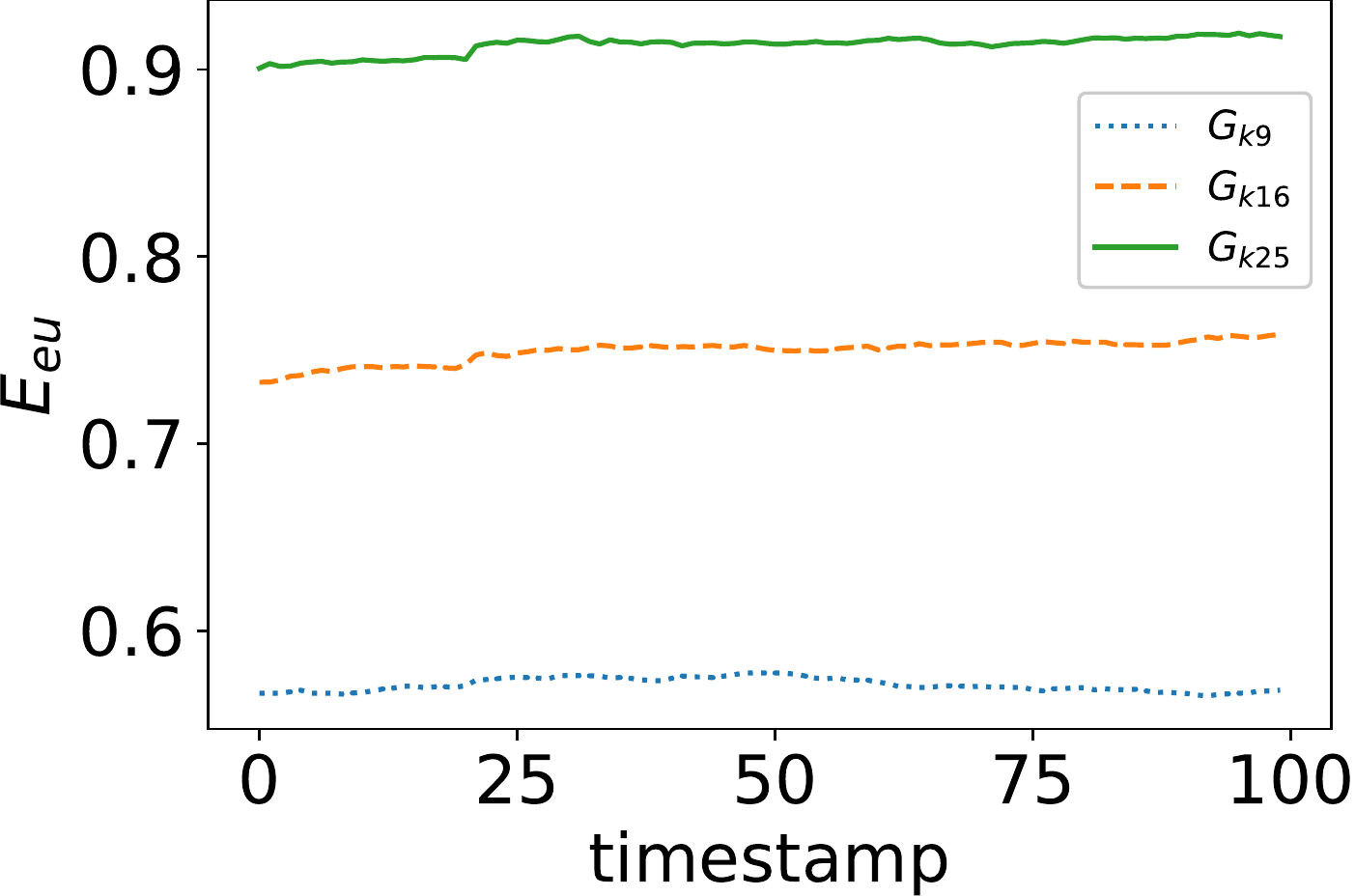}
	\end{minipage}
	\begin{minipage}{0.35\hsize}
		\includegraphics[width=\hsize]{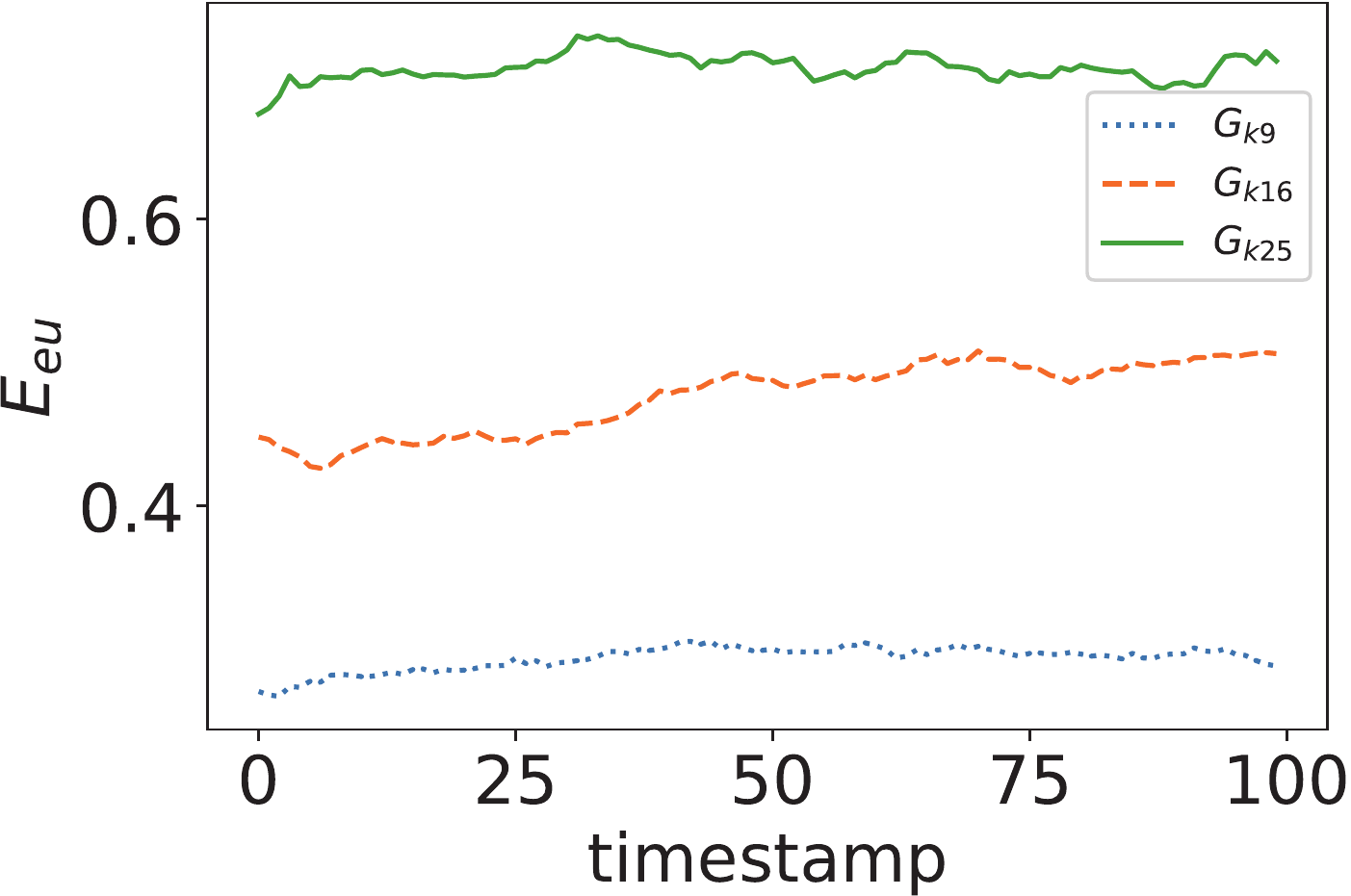}
	\end{minipage}
	\vspace{-5pt}
	\caption{\label{fig-traj2}{\small Utility of Private Trajectory Release with different Policy Graphs.}}
	\vspace{-20pt}
\end{figure}

\noindent \textbf{Evaluation of Location Trace Release.}
We demonstrate the utility of private trajectory release with PGLP in Fig. \ref{fig-traj1} and Fig.\ref{fig-traj2}.
In Fig. \ref{fig-traj1}, we show the results of P-LM and P-PIM on the Geolife Dataset.
We test  20 users' trajectories with 100 timestamps and report average  $ E_{eu} $ at each timestamp. 
We can see P-PIM has higher utility than P-PIM, which in line with the results for single location release.
The error $ E_{eu} $ is increasing along with timestamps due to the enlarged constrained domain, which is in line with Fig.\ref{fig:repair_graph_alg5_vs_base}.
The average of  $ E_{eu}$  across 100 timestamps on different policy graphs, i.e., $ G_{k9} $, $ G_{k16} $ and $ G_{k25} $ is also in accordance with the single location release in Fig.\ref{fig:util_on_graphs}.
$ G_{k9} $ has the least average error of $ E_{eu}$  due to the smallest sensitivity.

In Fig.\ref{fig-traj2}, we show the utility of P-PIM with different policy graphs on two different datasets  Geolife and  Gowalla.
The utility of $ G_{k9} $ is always the best over different timestamps for both datasets.
In general, the Gowalla dataset has better utility than the Geolife dataset because the constraint domain of the Gowalla dataset is smaller.
The reason is that the Gowalla dataset collects check-in locations that have an apparent mobility pattern, as shown in \cite{KDD-Gowalla-2011}.
While Geolife dataset collects  GPS trajectory with diverse transportation modes such as walk, bus, or train; thus, the trained Markov model is less accurate.

\vspace{-12pt}

\section{Conclusion}
\vspace{-8pt}
In this paper, we proposed a flexible and rigorous location privacy framework named PGLP, to release private location continuously under the real-world constraints with customized location policy graphs. 
We design an end-to-end private location trace release algorithm satisfying a pre-defined location privacy policy. 

For future work, there are several promising directions.
One  is to study how to use the rich interface of PGLP  for the utility-privacy tradeoff in the real-world location-based applications, such as carefully designing location privacy policies for COVID-19 contact tracing \cite{cao_panda:_2020}.
Another exciting direction is to design advanced mechanisms to achieve  location privacy policies with less noise.

	\bibliographystyle{abbrv}
{
	\bibliography{ref}}



\vspace{-10pt}
\begin{subappendices}
\renewcommand{\thesection}{\Alph{section}}%

\vspace{-10pt}
\section{An example of Isolated Node}
\label{appx-iso}
\vspace{-5pt}
\noindent{\bf Intuition.} 
We examine the privacy guarantee of P-PIM w.r.t. $ \mathcal{G}^\mathcal{C} $ in Fig.\ref{fig-ex-exposure}(a).
According to K-norm Mechanism \cite{Geometry-Hardt-STOC10} in Definition \ref{def-knorm}, P-PIM guarantees that, for any two neighbors  $\textbf{s}_{i}$ and $\textbf{s}_{j}$, their difference is bounded in the convex body {\small $K$}, i.e. {\small $f(\textbf{s}_{i})-f(\textbf{s}_{j})\in K$}. 
Geometrically, for a location $\textbf{s} $, all other locations  in the convex body of {\small $K+f(\textbf{s})$} are $\epsilon$-indistinguishable with $\textbf{s}$. 

\begin{figure}[H]
	\centering
	\vspace{-40pt}
	\includegraphics[width=6cm]{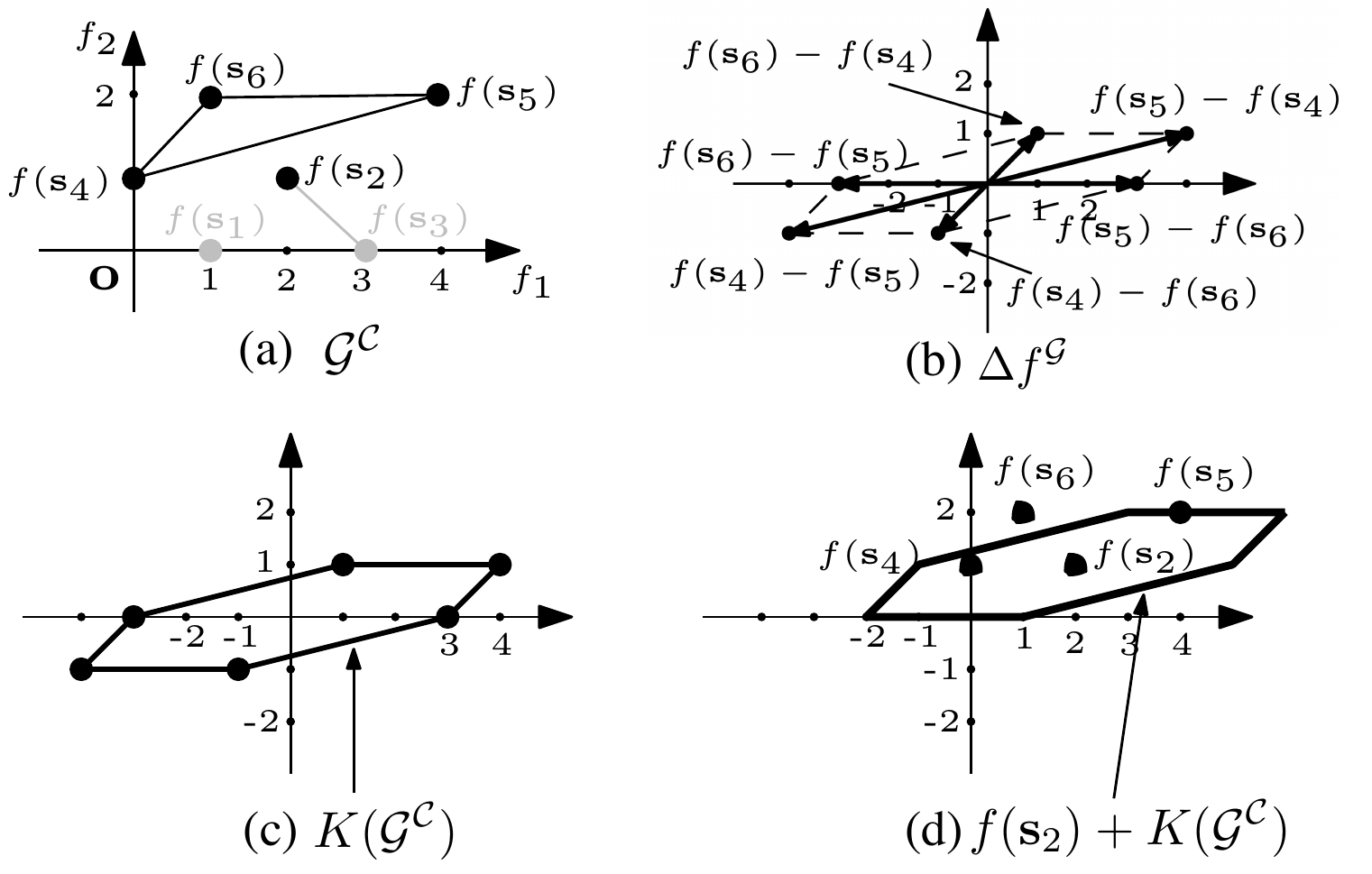}
	\vspace{-10pt}
	\caption{{\small (a) A policy graph under {\footnotesize $\mathcal{C}=\{\textbf{s}_2,\textbf{s}_4,\textbf{s}_5,\textbf{s}_6\}$}; 
			(b) the {\small $\Delta f^\mathcal{G}$ }of vectors {\footnotesize $f(\textbf{s}_{i})-f(\textbf{s}_{j})$}; 
			(c) the sensitivity hull {\footnotesize $ K({\mathcal{G}^\mathcal{C}}) $} covering the {\small $\Delta f^\mathcal{G}$}; 
			(d) the shape {\footnotesize $f(\textbf{s}_{2})+K({\mathcal{G}^\mathcal{C}})$} containing $f(\textbf{s}_{4})$ and $f(\textbf{s}_{5})$. 
			That is to say, $ \textbf{s}_2 $ is indistinguishable with $ \textbf{s}_4 $ and $ \textbf{s}_5$.
		}
	}
	\label{Figure-not-isolated}
	\vspace{-20pt}
\end{figure}

\vspace{-5pt}
\begin{example}[Disconnected but Not Isolated Node]
	\label{example-dis-iso}
	In Figure \ref{Figure-not-isolated}, $\textbf{s}_2$ is disconnected under constraint {\small $\mathcal{C}=\{\textbf{s}_2,\textbf{s}_4,\textbf{s}_5,\textbf{s}_6\}$}. However, $\textbf{s}_{2}$ is not isolated because {\small $f(\textbf{s}_2)+K$} contains {\small $f(\textbf{s}_4)$ }and {\small $f(\textbf{s}_5)$}. Hence, {\small $\textbf{s}_{2}$} and other nodes in $\mathcal{C}$ are indistinguishable.
\end{example}

\vspace{-15pt}

\section{Policy Graph Repair Algorithm}
\label{appx-repair}

\vspace{-5pt}

	Figure \ref{Figure-Gt-example}(a) shows the graph under constraint {\small $\mathcal{C}=\{\textbf{s}_3,\textbf{s}_4,\textbf{s}_5,\textbf{s}_6\}$}. Then $\textbf{s}_3$ is exposed because {\small $f(\textbf{s}_3)+K^\mathcal{G}$} contains no other node. 
	To satisfy the PGLP without isolated nodes, we need to connect $\textbf{s}_3$ to
	another node in $\mathcal{C}$, i.e. $\textbf{s}_4$, $\textbf{s}_5$ or $\textbf{s}_6$.
	
	If $\textbf{s}_3$ is connected to $\textbf{s}_4$, then Figure \ref{Figure-Gt-example}(b) shows the new graph and its sensitivity hull. By adding two new edges {\scriptsize $\{f(\textbf{s}_3)-f(\textbf{s}_4),f(\textbf{s}_4)-f(\textbf{s}_3)\}$} to $\Delta f^\mathcal{G}$, the shaded areas are attached to the sensitivity hull. Similarly, Figures \ref{Figure-Gt-example}(c) and (d) show the new sensitivity hulls when $\textbf{s}_3$ is connected to $\textbf{s}_6$ and $\textbf{s}_5$ respectively. Because the smallest area of {\small $K^\mathcal{G}$} is in Figure \ref{Figure-Gt-example}(b), the repaired graph is $\mathcal{G}^\mathcal{C}\lor \overline{\textbf{s}_3\textbf{s}_4}$, i.e., add adge $  \overline{\textbf{s}_3\textbf{s}_4}$ to the graph  $\mathcal{G}^\mathcal{C}$.

\begin{theorem}
	Algorithm \ref{alg-repair} takes $O(m^2log(m))$ time where $ m $ is the number of valid nodes (with edge) in the policy graph.
\end{theorem}

\noindent{\bf Algorithm.}
Algorithm \ref{alg-repair} derives the minimum protectable graph for location data when an isolated node is detected. We can connect the isolated node $\textbf{s}_i$ to the rest (at most $m$) nodes, generating at most $m$ convex hulls  where $m=|\mathcal{V}|$ is the number of valid nodes. 
In two-dimensional space, it only takes $O(mlog(m))$ time to find a convex hull.  
To derive ${Area}$ of a shape, we exploit the computation of determinant whose intrinsic meaning is the $\textsc{Volume}$ of the column vectors. Therefore,
we use {\scriptsize $\mathop\sum\limits_{i=1,j=i+1}^{i=h}det(\textbf{v}_i,\textbf{v}_j)$} to derive the {Area} of a convex hull with clockwise nodes $\textbf{v}_1,\textbf{v}_2,\cdots,\textbf{v}_{h}$ where $h$ is the number of vertices and $\textbf{v}_{h+1}=\textbf{v}_1$.
By comparing the area of these convex hulls, we can find the smallest area in $O(m^2log(m))$ time where $m$ is the number of valid nodes.
Note that Algorithms \ref{alg-exposure} and \ref{alg-repair} can also be combined together to protect any disconnected nodes.

\vspace{-10pt}

\section{Related Works}
\vspace{-10pt}

\subsection{Differential Privacy}
%
While differential privacy \cite{Dwork06differentialprivacy}  has been accepted as a standard notion for privacy protection,
 the concept of standard differential privacy is not generally applicable for complicated  data types.
Many variants of differential privacy have been proposed, such as  Pufferfish privacy \cite{Kifer-2012-pufferfish}, Geo-Indistinguishability \cite{geoi_ccs13} and Voice-Indistinguishability \cite{han_voice-indistinguishability:_2020} (see  Survey \cite{pejo_sok:_2020}).
Blowfish privacy \cite{Blowfish-SIGMOD14} is the first generic framework with customizable privacy policy. 
It defines sensitive information as secrets and known knowledge about the data as constraints. 
By constructing a policy graph, which should also be consistent with all constraints, Blowfish privacy can be formally defined. 
Our definition of PGLP is inspired by Blowfish framework. 
Notably, we find that the policy graph may not be viable under temporal correlations represented by Markov model, which was not considered in the previous work.
This is also related to another line of works studying how to achieve differential privacy under temporal correlations  \cite{cao_contpl:_2018, cao_quantifying_2017, cao_quantifying_2019, song_pufferfish_2017}.

\vspace{-10pt}
\subsection{Location Privacy}
A close line of works focus on extending differential privacy to location setting.
The first DP-based location privacy model is Geo-Indistinguishability\cite{geoi_ccs13}, which scales the sensitivity of two locations to their Euclidean distance. 
Hence, it is suitable for proximity-based applications.
Following by Geo-Indistinguishability, several location privacy notions \cite{xiao_loclok:_2017, takagi_GGI_2019}  have been proposed based on differential privacy.
A recent  DP-based location privacy,  spatiotemporal event privacy\cite{cao_priste:_2019, cao_priste:_2019-1, cao_protecting_2019},  proposed a new representation for secrets in spatiotemporal data called spatiotemoral events using Boolean expression.
It is essentially different from this work since here we are considering the traditional representation of secrets, i.e., each single location or a sequence of locations.

Several works considered Markov models for improving utility of released location traces or web browsing activities \cite{ChatzikokolakisPS14,liyue-2014-www}, but did not consider the inference risks when an adversary has the knowledge of the Markov model.
Xiao et al. \cite{xiao_ccs15} studied how to protect the true location if a user's movement follows Markov model. 
The technique can be viewed as a special instantiation of PGLP.
In addition, PGLP uses a policy graph to tune the privacy and utility to meet diverse the requirement of  location-based applications.

\end{subappendices}

\end{document}